\documentclass{aa}

\usepackage{graphicx,amsmath}
\usepackage{natbib}
\usepackage{bm}

\usepackage{arydshln}

\usepackage{txfonts}
\usepackage{color}

\begin{document}

   \title{GALACTICNUCLEUS: A high-angular-resolution $JHK_s$ imaging survey of the Galactic centre}
   \titlerunning{GALACTICNUCLEUS}
   \authorrunning{Nogueras-Lara et al.}

   \subtitle{II. First data release of the catalogue and the most detailed CMDs of the GC\thanks{The catalogues described in this paper and specified in Table \ref{cat} are available in electronic form at the CDS via anonymous ftp to cdsarc.u-strasbg.fr (130.79.128.5) or via http://cdsweb.u-strasbg.fr/cgi-bin/qcat?J/A+A/. The catalogues  follow the structure presented in Table \ref{final_cat}.}}

  \author{F. Nogueras-Lara
          \inst{1}
          \and      
          R. Sch\"odel 
          \inst{1}       
          \and
          A. T. Gallego-Calvente
          \inst{1}
          \and
         H. Dong
          \inst{1}
          \and
           E. Gallego-Cano
          \inst{1,2}
          \and
           B. Shahzamanian
          \inst{1}
          \and          
          J.~H.~V. Girard
          \inst{3}
         \and
          S. Nishiyama
          \inst{4}
         \and
          F. Najarro 
          \inst{5}   
         \and
          N. Neumayer 
          \inst{6}             
          }

   \institute{
    Instituto de Astrof\'isica de Andaluc\'ia (CSIC),
     Glorieta de la Astronom\'ia s/n, 18008 Granada, Spain
              \email{fnoguer@iaa.es}
         \and
              Centro Astron\'omico Hispano-Alem\'an (CSIC-MPG), Observatorio Astron\'omico de Calar Alto, Sierra de los Filabres, 04550, G\'ergal, Almer\'ia, Spain     
          \and
    Space Telescope Science Institute, Baltimore, MD 21218, USA
         \and
      Miyagi University of Education, Aoba-ku, 980-0845 Sendai, Japan
      \and
       Departamento de Astrof\'isica, Centro de Astrobiolog\'ia (CSIC-INTA), Ctra. Torrej\'on a Ajalvir km 4, E-28850 Torrej\'on de Ardoz, Spain
       \and 
       Max-Planck Institute for Astronomy, K\"onigstuhl 17, 69117 Heidelberg, Germany
           }
   \date{}

 
  \abstract
   {The high extinction and extreme source crowding of the central regions of the Milky Way are serious obstacles to the study of the structure and stellar population of the Galactic centre (GC).  Existing surveys that cover the GC region (2MASS, UKIDSS, VVV, SIRIUS) do not have the necessary high angular resolution. Therefore, a high-angular-resolution survey in the near infrared is crucial to improve the state of the art.}
   {Here, we present the GALACTICNUCLEUS catalogue, a near infrared $JHK_s$ high-angular-resolution ($0.2''$) survey of the nuclear bulge of the Milky Way.}
   {We explain in detail the data reduction, data analysis, calibration, and uncertainty estimation of the GALACTICNUCLEUS survey. We assess the data quality comparing our results with previous surveys. 
}
{We obtained accurate $JHK_s$ photometry for $\sim 3.3\times10^6$ stars in the GC detecting around 20 \% in $J$, 65 \% in $H,$ and 90 \% in $K_s$. The survey covers a total area of $\sim0.3$ square degrees, which corresponds to $\sim 6,000$ pc$^2$. The GALACTICNUCLEUS survey reaches 5\,$\sigma$ detections for $J \sim 22$ mag, $H \sim 21$ mag, and $K_s \sim 21$ mag. The uncertainties are below 0.05 mag at $J \sim 21$ mag, $H \sim 19$ mag, and $K_s \sim 18$ mag. The zero point systematic uncertainty is $\lesssim0.04$ mag in all three bands. We present colour-magnitude diagrams for the different regions covered by the survey.}
   
   {}


\titlerunning{GALACTICNUCLEUS. II.}
\authorrunning{F. Nogueras-Lara et al.}

   \maketitle
%

\section{Introduction}

The Galactic centre (GC) is a major astrophysical target since it is the closest galactic nucleus located at only 8 kpc from Earth, about a hundred times closer than the Andromeda galaxy, and a thousand times closer than the next active galactic nucleus. The GC hosts a supermassive black hole (Sgr A*) located at the dynamical centre of the Galaxy and surrounded by the nuclear star cluster, almost ten times more massive than Sgr A* \citep{Launhardt:2002nx,Schodel:2014fk} and composed of a complex stellar population with the majority of the stars being older than 5 Gyr \citep{Pfuhl:2011uq}. On a larger scale, the NSC and Sgr A* are embedded in, and form part of, the nuclear bulge (NB). The NB is a distinct stellar structure that stands clearly out from the kiloparsec-scale Galactic Bulge/Bar and constitutes a flattened, possibly disc-like structure -nuclear stellar disc (NSD)- with a radius of $\sim$230 pc and a scale height of $\sim$45 pc \citep{Launhardt:2002nx}.

The GC is characterised by the most extreme conditions
in the Galaxy: extreme stellar densities \citep[$\sim 10^{5-7}$ pc$^{-3}$,][]{Launhardt:2002nx,Schodel:2007tw,Schodel:2018aa}, a tidal field so intense that even massive, young clusters dissolve into the background in less than 10 Myr \citep{Portegies-Zwart:2002fk}, high turbulence and temperature of the interstellar medium \citep{Morris:1996vn}, a strong magnetic field \citep{Crocker:2012fk}, and intense UV radiation \citep{Launhardt:2002nx}. Despite, or possibly because of these extreme properties, the GC is the most prolific massive star forming environment in the Galaxy \citep{Schodel:2007tw,Yusef-Zadeh:2009ph,Mauerhan:2010kb}.

It cannot be stressed enough that the centre of the Milky Way is the
only galaxy nucleus in which we can actually resolve the NSC and the NB observationally and examine its properties and dynamics. Nevertheless, only small regions like the central parsec and the Arches and Quintuplet clusters, which represent around 1\% of the projected area, have been explored in detail. 

To characterise the GC stellar population it is necessary  to overcome the high stellar crowding and the extreme interstellar extinction \citep[$A_V\gtrsim30$, $A_{K_{s}}\gtrsim2.5$, e.g.][]{Scoville:2003la,Nishiyama:2008qa,Fritz:2011fk,Schodel:2010fk,Nogueras-Lara:2018aa}. This requires an angular resolution of $\sim0.2''$ and multi-band observations. Several large imaging surveys include the GC region (e.g. 2MASS, UKIDSS, VVV, SIRIUS/IRSF), but they are limited in angular resolution to $>0.6''$ by atmospheric seeing. Hence, their photometry is inaccurate and their completeness limit is as shallow as $K\sim14$ mag. Additionally, stars brighter than $K \approx 9-10$ mag are usually heavily saturated in these surveys.

In this paper we present the first data release of the GALACTICNUCLEUS survey,  a high-angular-resolution $\sim 0.2''$, multi-wavelength ($J$, $H$ and $K_s$) imaging survey especially designed to observe the GC. This constitutes the first survey of the central regions of our Galaxy with $\sim 10$ mag dynamic range in three bands in the near infrared (NIR). 
Several papers that use the GALACTICNUCLEUS survey and show its potential have already been published \citep{Nogueras-Lara:2018aa,Nogueras-Lara:2018ab} or submitted (Gallego-Cano et al., submitted; Nogueras-Lara et al., submitted).

This paper constitutes the second paper of a series initiated with \citet{Nogueras-Lara:2018aa}.

\section{Observations}

The GALACTICNUCLEUS survey consists of 49 pointings toward the GC and the inner Bulge (Fig\,\ref{scheme_white}). The observations were carried out with HAWK-I \citep[High Wide field K-band Imager, ][]{Kissler-Patig:2008fr} located at the ESO VLT unit telescope 4 \footnote{Based on observations made with ESO Telescopes at the La Silla Paranal Observatory under programme ID 195.B-0283}. We used the broadband filters $J$, $H$ and $K_s$ to cover the NIR regime.  HAWK-I has four HAWAII 2RG 2048 $\times$ 2048 detectors (four chips) with a cross-shaped gap of 15'' between them. The on-sky field of view of HAWK-I is 7.5' $\times$ 7.5' and its pixel scale is 0.106$''$/pixel. We used the fast photometry mode to obtain a series of short-exposure frames with an exposure time of $DIT = 1.26\,s$ (detector integration time) that allowed us to improve the angular resolution of the final images to $\sim 0.2''$ through applying the speckle holography algorithm \citep{Schodel:2013fk}. Due to the short readout time, we had to window the detector, which resulted in a field of view (FoV) of $2048\times768$ pixels for each of the four chips. Tables \ref{obs_cen}, \ref{obs_bul}, and \ref{obs_disc} summarise the observing conditions for each pointing.

\subsection{Observing strategy}

We treated each of the four HAWK-I detectors in a completely independent way. To cover the gap between the detectors and to achieve some overlap between the pointings, we applied random jittering with a jitter box varying in width from $30''$ (2015 data) to $1'$ (2016-2018 data). The jitter box was increased after the first epoch to optimise the data coverage in the cross-shaped gap.

\begin{figure}
   \includegraphics[width=\linewidth]{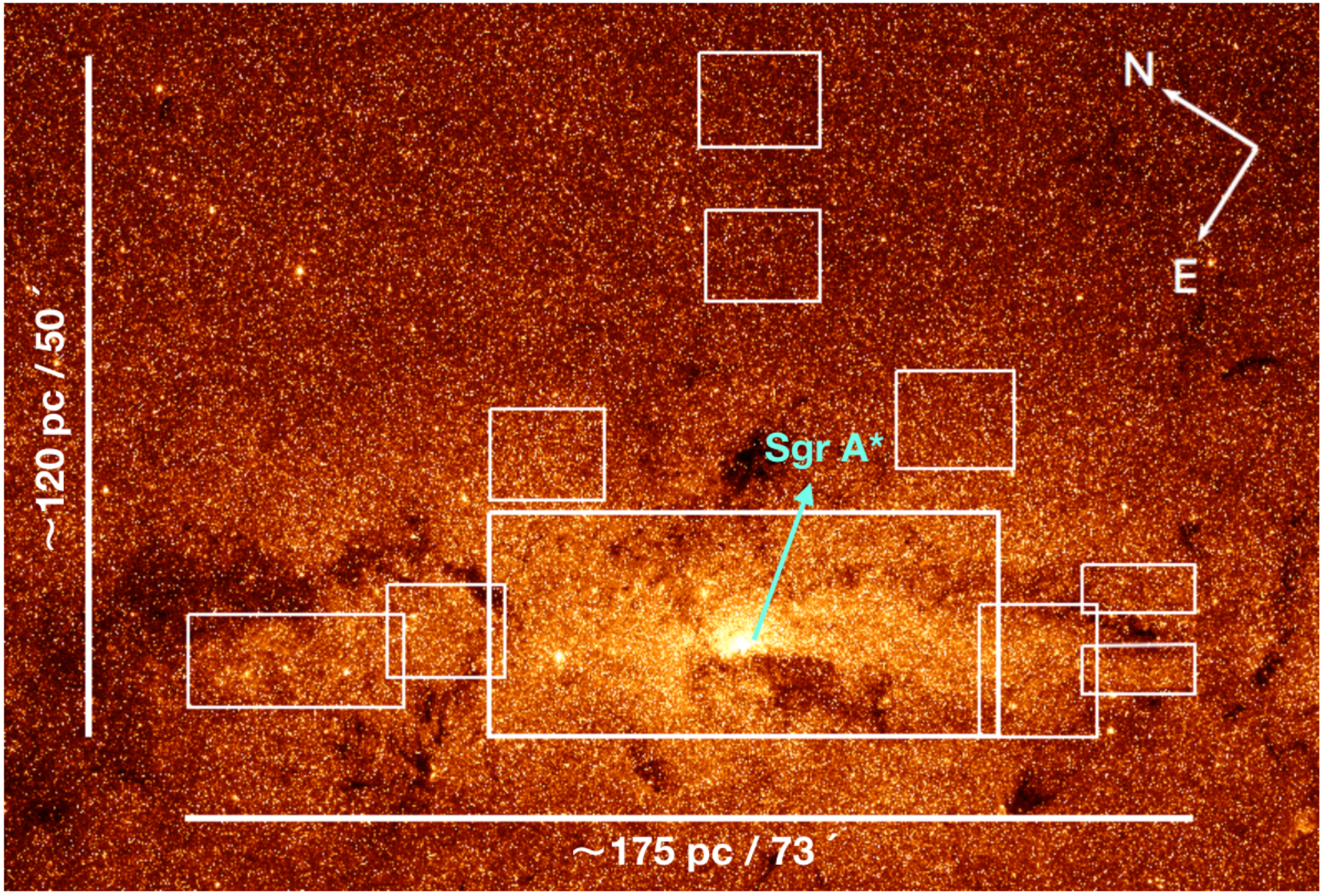}
   \caption{Scheme of the target fields for the GALACTICNUCLEUS survey over-plotted on a Spitzer/IRAC image at 3.6 $\mu$m. The position of Sagittarius A* is highlighted in cyan.}

   \label{scheme_white}
    \end{figure}

\begin{figure*}
   \includegraphics[width=\linewidth]{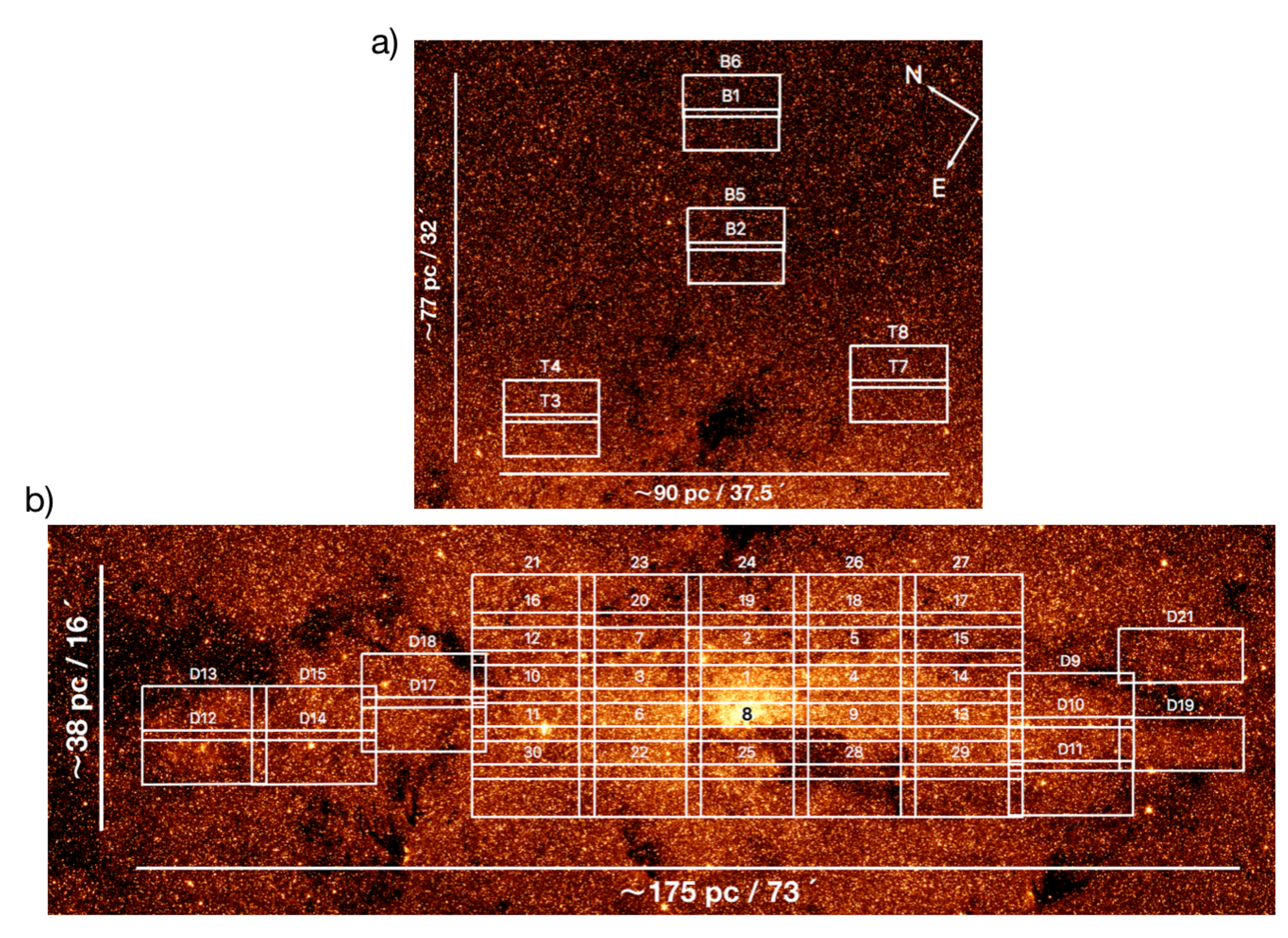}
   \caption{Detailed scheme of all the fields observed in the GALACTICNUCLEUS survey. Each white square represents a field of the survey. Labels are included in white to identify the fields. Fields in the NSD, inner Galactic bulge, and the transition zone between the bulge and the centre are identified by the letters `D', `B' and `T'. Fields in the central region are identified by numbers from 1 to 30.}
   
   \label{scheme_zoom_white}
    \end{figure*}

The 49 pointings cover the GC in four distinct groups: (1) A continuous, rectangular area of about $36'\times16'$ centred on Sgr A* (30 pointings, labelled 1-30), (2) eleven pointings (D9-D15, D17-D19 and D21) toward the east and west of the central field to cover low-extinction areas in the central part of the nuclear stellar disc, (3) four pointings (T3, T4, T7 and T8) in the transition zone between the NB and the inner bulge, and (4) four pointings (B1, B2, B5 and B6) toward comparison fields in the inner bulge just north of the nuclear disc. The entire survey area is outlined in Fig.\,\ref{scheme_white}, while Fig.\,\ref{scheme_zoom_white} allows identification of each pointing. All pointings overlap with at least one adjacent pointing. In this way, we were able to compare the common stars to assess the data quality.

We rotated HAWK-I 31.40$^\circ$ east of north to align the observed fields with the Galactic Plane \citep[][]{Reid:2004ph}.  Due to the extreme source density toward the GC, simple jittering on target will not work to obtain accurate measurements of the sky background. Instead, we chose the following strategy:  Sky frames were taken just before or after each science observation on a dark cloud with very low stellar density located at approximately 17$^h$ 48$^m$ 01.55$^s$, -28$^\circ$ 59$'$ 20$''$. We rotated the camera 70$^\circ$ east of north to align the rectangular FoV with the extension of the dark cloud. From the sky observations we created a master sky frame. We then scaled the master sky to the instantaneous sky background of each exposure, which was estimated from the median value of the 10\% of pixels with the lowest values. A detector dark image was subtracted from both the master sky and from each reduced science frame before determining this scaling factor \citep[see also ][for a detailed description of the methodology]{Nogueras-Lara:2018aa}.

\section{Data reduction and analysis}
\label{summary}

\begin{figure}
   \includegraphics[width=\linewidth]{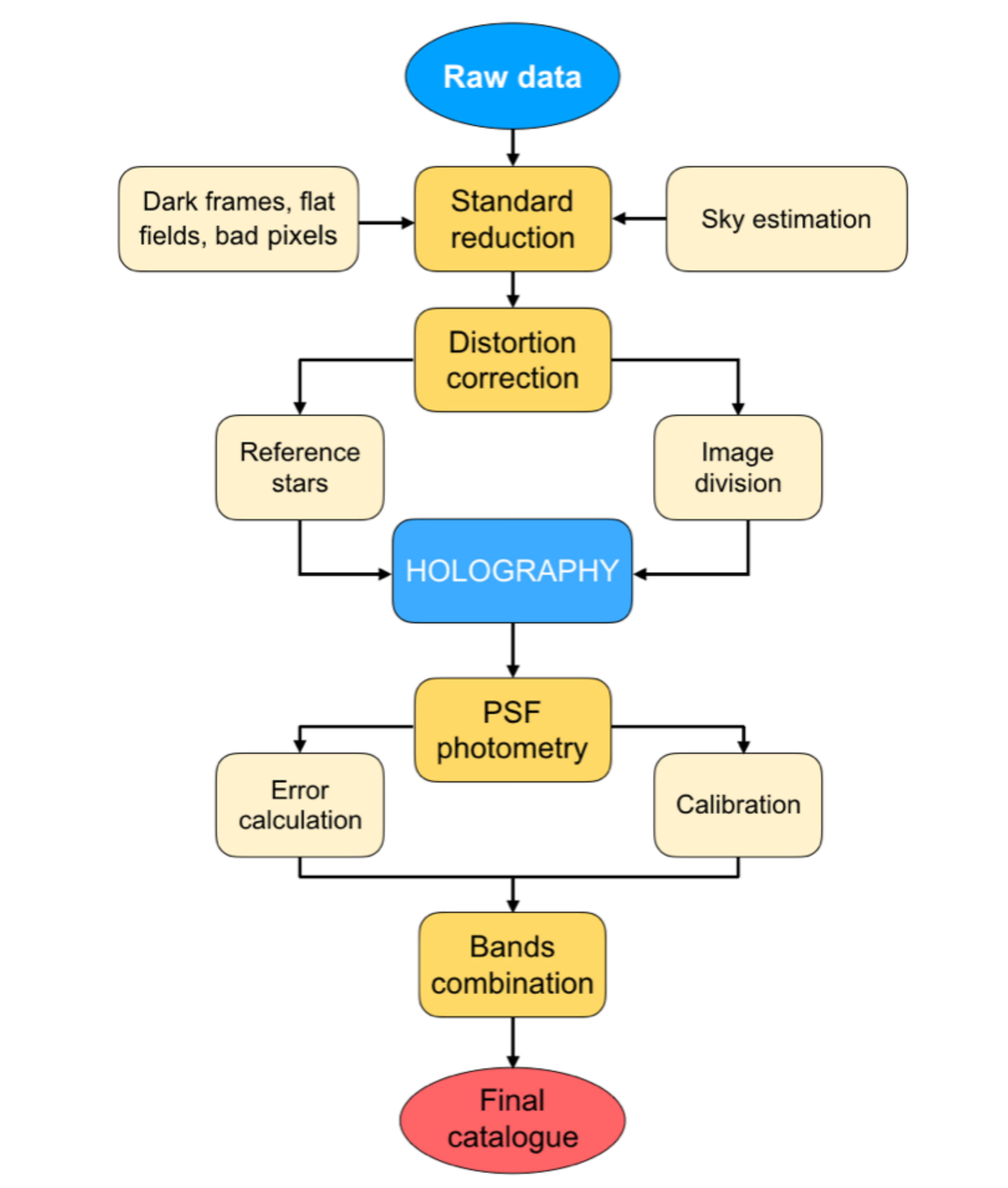}
   \caption{Scheme of the obtention of the final catalogue for each chip.}
   
   \label{scheme_reduction}
    \end{figure}

The reduction and treatment of the central pointing of the survey (F1) was described in detail in \citet{Nogueras-Lara:2018aa}. The same procedure was used for the whole survey. Briefly summarised, it consists of the following steps: 
\begin{enumerate}
\item Standard reduction: Dark subtraction, flat fielding, sky subtraction and bad pixel correction. We also aligned the frames correcting for the jittering. 
\item Distortion correction: we used stellar positions from  VVV $J$-band images to compute a geometric distortion solution for each of the bands and fields observed. 
\item Speckle holography: We used the speckle holography method \citep[see e.g.][]{Primot:1990fk,Petr:1998vn} optimised for crowded fields \citep{Schodel:2013fk}. This is an image reconstruction technique that combines a high number of short-exposure frames ($\sim 1000$ per pointing in our case) using an averaged division of quantities in Fourier space \citep[eq. 1,][]{Schodel:2013fk}. The final product is convolved with a Gaussian of  full width at half
maximum (FWHM) of 0.2$''$. This determines the final angular resolution of the catalogue. Since this technique requires the knowledge of the instantaneous PSF of each exposure, which can vary spatially due to anisoplanatic effects, we divided each frame into small subregions and used the reference stars in each of them to extract local, instantaneous PSFs \citep[see detailed description in Sect.\,2.3 of][]{Nogueras-Lara:2018aa}. 
\item Photometry: We used the {\it StarFinder} software package \citep{Diolaiti:2000qo} to perform PSF photometry and astrometry for each chip independently using the final holographic product. Since {\it StarFinder} underestimates the uncertainties computed for individual stars (Emiliano Diolaiti, private communication), we used an alternative approach. We generated three independent holographic images for each chip (sub-images) using one third of the data for each of them. We also created a deep final image using all the data available. We extracted PSF photometry from each sub-image and from the deep one. We accepted a star only if it was detected in all three sub-images. We calculated the final flux of each star using the deep image and the uncertainty by means of the equation \citep{Nogueras-Lara:2018aa}:

\begin{equation}
\label{eq}
\Delta f = \frac{f_{max}-f_{min}}{2 \sqrt{N}} ,
\end{equation}
\noindent where $f_{max}$ and $f_{min}$ are the maximum and minimum fluxes obtained for each star in the sub-images and $N=3$ is the number of sub-images.

This strategy is quite conservative, but it allows us to keep only real detections and to estimate the uncertainties based on three independent data sets. We also considered the possible variation of the PSF across the detector. We divided each chip into three equal regions from which we extracted three independent PSFs. We estimated the uncertainty associated to its variation by comparing the PSFs computed for each region \citep[for details, see Sect.\,3.1.2 of ][]{Nogueras-Lara:2018aa}. For all the 49 pointings and each individual chip, this PSF uncertainty is $\lesssim 0.025$ mag in all three bands. We added this uncertainty quadratically  to the statistical uncertainty that was determined from the three sub-images. 
\item Calibration: We computed the astrometric solution by using common stars between each chip and the VVV catalogue. We estimated the uncertainty of this procedure to $\lesssim0.05$\,arcseconds. The photometric calibration was carried out with stars common with the SIRIUS/IRSF GC survey \citep[e.g.][]{Nagayama:2003fk,Nishiyama:2006tx}, which was specifically designed  for the study of the GC. We used bright stars (non-saturated) with low uncertainties $<0.05$ mag in all three bands in our catalogue. 
\item Field combination: We obtained the final list of stars for each filter and pointing by combining the stars detected on each chip. We corrected the small shifts in magnitude that can appear between the chips by using stars in the overlap region of all four of them and finding the factor for each chip that minimised the overall $\chi^2$ \citep[for further details, see ][]{Dong:2011ff,Nogueras-Lara:2018aa}. We then combined the star lists of the four chips. For stars detected in more than one chip we computed their photometry as the mean of the detections and their uncertainties as the quadratic combination of each uncertainty. After this procedure, we recalibrated the final star list of each pointing with the SIRIUS/IRSF GC survey  to avoid any photometric offsets that might have been introduced during the process. Finally, we combined all three bands and generated a catalogue for each field. Figure \ref{scheme_reduction} shows a scheme of the whole process. 
\end{enumerate}

\section{Preparation of the catalogues}

We prepared several catalogues: 1) Four catalogues for the regions (consisting of two pointings each) shown in Fig.\,\ref{scheme_zoom_white}a). They correspond to the inner bulge and the transition region between the inner bulge and the NB. 2) Three catalogues for the central region shown in Fig. \ref{scheme_zoom_white}b): GC West (pointings D12 to D18), GC Central (pointings 1 to 30), and GC East (pointings D9 to D11, D19 and D21).  

\def\arraystretch{1.25}
\begin{table}
\begin{center}
\caption{Fields included in each catalogue.}
\label{cat} 
\begin{tabular}{cc}
 & \tabularnewline
\hline 
\hline 
Catalogue & Fields included\tabularnewline
\hline 
Central & 1-30\tabularnewline
NSD East & D12-D18\tabularnewline
NSD West & D9-D11, D19, D21\tabularnewline
Transition East & T3-T4\tabularnewline
Transition West & T7-T8\tabularnewline
Inner Bulge North & B1,B6\tabularnewline
Inner Bulge South & B2,B5\tabularnewline
\hline 
\end{tabular}
\end{center}
\vspace{0.75cm}

\textbf{Notes.} The name of each field starts by a capital letter that indicates its position in the survey: `D' -> Nuclear stellar disc, `B' -> Inner bulge, and `T'-> transition zone inner bulge-nuclear stellar disc. Fields in the central part of the survey are simply numbered 1 to 30.

 \end{table}

\subsection{Astrometry}
\label{astrometry}
\begin{figure*}
   \includegraphics[width=\linewidth]{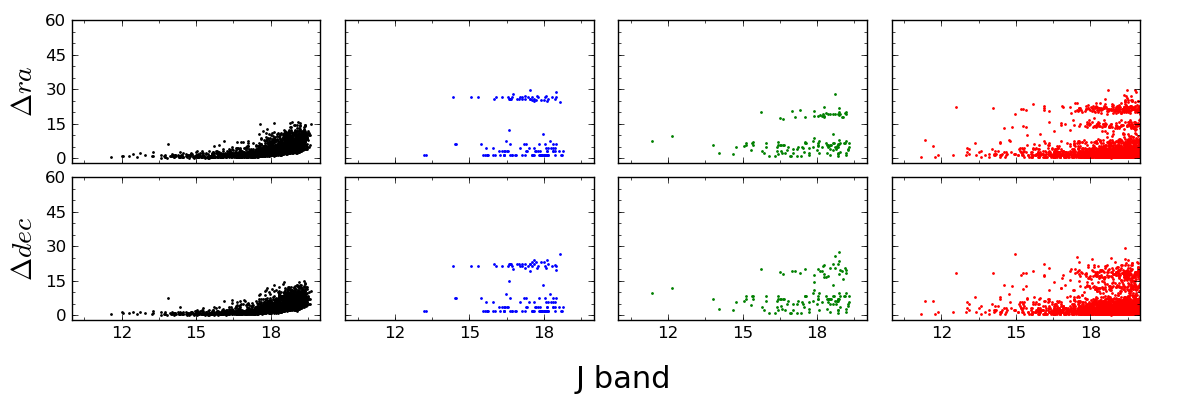}
   \caption{Analysis of the astrometric uncertainties for the $J$-band. The y-axes is in units of milli-arcseconds. Each column depicts the uncertainties associated to: 1) relative astrometry measured on a single chip and pointing (F6, chip1 is shown as an example); 2) a combination of chips for a given pointing (F6 shown as an example); 3) a combination of all fields for the central region; 4) a combination of the three different bands for the central region. The plots in columns 2, 3, and 4 show steps in the distributions which are caused by systematic offsets between different chips, pointings, and filters.}
   
   \label{pos_uncer}
    \end{figure*}

For each band, we identified common stars in the overlapping regions and computed their positions through averaging the multiple detections. The final catalogue was created by combining all three bands. The  maximum allowed offset for a star present in different bands was set to $\sim0.1''$, half the angular resolution of the images. The final positions of the stars detected in more than one band were computed as the mean positions of the individual detections. The absolute astrometric uncertainty is derived from the alignment with the VVV stars that were used to compute the astrometric solution. We estimated an upper limit of the absolute astrometric uncertainty  of $0.05 ''$ with respect to VVV. As concerns the relative astrometric uncertainty of the stars, we examined the following sources of uncertainty: 1) uncertainties derived from the measurements in the three independent sub-images for each band and pointing (see Sect. \ref{summary}); 2) uncertainties measured from the common stars in the overlapping area between the four chips for each pointing; 3) uncertainties estimated from common stars when  combining the different pointings to produce the final list for each band; and 4) uncertainties estimated from  the combination of the positions of the stars from the three different bands. Figure\,\ref{pos_uncer} shows the corresponding uncertainties for chips, pointings, and the central catalogue for the case of the $J$-band. We concluded that the relative astrometric uncertainty is dominated by the combination of the different pointings and bands (points 3 and 4 above).


The final catalogues include the uncertainty due to (3) and (4) for the case of stars detected in more than one field and/or detected in more than one band. For the remaining stars we estimated an upper limit of $\sim$30 milliarcsec, which is the maximum value obtained in the analysis presented in Fig. \ref{pos_uncer}.

\subsection{Photometry}
\label{rec_zp}

\begin{figure}
   \includegraphics[width=\linewidth]{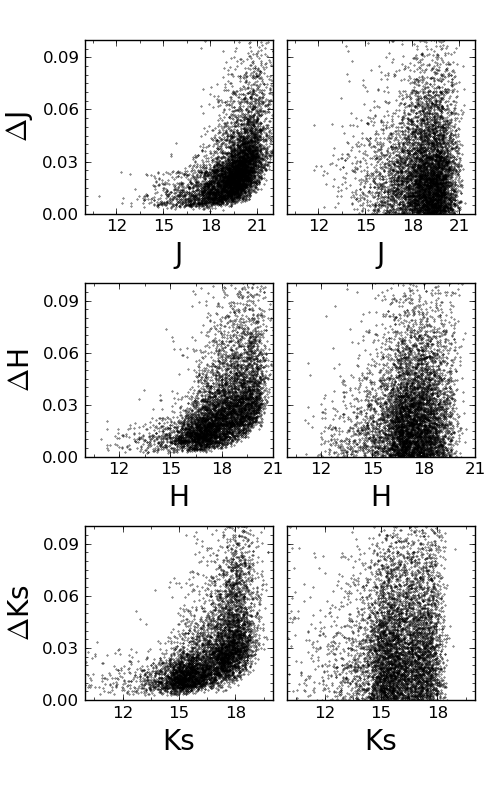}
   \caption{Left panels: Photometric uncertainties due to the quadratic propagation of the errors for each star. Right panels: Uncertainties computed for stars detected in the overlapping regions between fields using Eq. \ref{eq}. The uncertainties shown in the right panels reach lower magnitudes since those stars need to be detected in more than one field, and the slightly different atmospheric conditions between pointings complicate the detection of the faintest stars. Only a random fraction of the stars has been plotted for clarity.}
   
   \label{photometric_uncertainty}
    \end{figure}

For stars detected in more than one pointing we computed the mean flux as the average value of the individual detections and estimated the photometric uncertainty by means of the quadratic propagation of the individual uncertainties of each detection.  We compared the uncertainties derived from multiple detections with the one determined from the measurements in the three sub-images (Eq.\,\ref{eq}). We used a two-sigma clipping algorithm to calculate the mean of the uncertainties obtained with both methods and concluded that both values are similar for all three bands ($\sim 0.03$\,mag). Figure \ref{photometric_uncertainty} shows the uncertainties obtained using both methods. 

For the combination of the pointings, we recomputed the zero point (ZP) using the Sirius catalogue when adding new pointings to the final catalogue. In this way, we avoided small photometric shifts that can appear when computing the average values for common stars.

We ended up with an uncertainty $\lesssim 0.05$ mag for all three bands at $J\lesssim21$ mag, $H\lesssim19$ mag, and $K_s\lesssim18$ mag, which is slightly better than for the case of the F1 alone (the central most field that suffers the worst crowding) that was analysed in \citet{Nogueras-Lara:2018aa}.

\subsubsection{Calibration}

To calibrate GALACTICNUCLEUS, we used the NIR $JHK_s$ SIRIUS IRSF survey \citep[e.g.][]{Nagayama:2003fk,Nishiyama:2006tx}. This catalogue was specially designed to study the GC and it uses PSF photometry, which allows the photometry to be improved in crowded fields. Thus, its characteristics make this survey a very appropriate reference. Moreover, we compared both photometric systems to check whether there is any significant difference between the filters:

\begin{itemize}
\item We computed the effective wavelength obtained for a red clump  (RC) star located at the GC distance with an extinction of $A_{K_s} \sim 1.9$ mag, corresponding to the extinction expected at the GC \citep{Nogueras-Lara:2018aa}. We obtained that the effective wavelength for the equivalent filters differs by $\lesssim 0.5\%$.\\

\item We also calculated the photometry in both photometric systems for a RC star and an early type star (T = 45,000 K) located at the GC distance and assuming an extinction of $A_{K_s} \sim 1.9$ mag. We obtained that the difference is $\lesssim 0.8\%$ in all three bands.\\

\end{itemize}

Therefore, we concluded that the differences between both photometric systems are negligible.

\subsubsection{Zero point}

\begin{figure}
   \includegraphics[width=\linewidth]{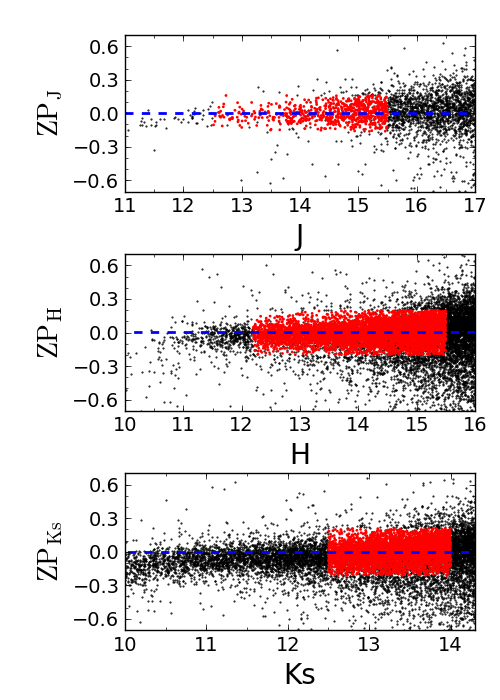}
   \caption{Comparison of the photometry of GALACTICNUCLEUS and  SIRIUS stars after obtaining the final lists for each band in the central catalogue. All common stars between SIRIUS and GALACTICNUCLEUS are plotted in black. Red points indicate those used for calibration. The deviation for bright stars is caused by  saturation.}
   
   \label{ZP_second}
    \end{figure}

Once the final lists had been  produced for each band, we recomputed the ZP  again using the SIRIUS/IRSF GC survey. This calibration was done in the same way as explained in \citet{Nogueras-Lara:2018aa}. We identified non-saturated common stars (coincident in position within a radius of $\sim 0.1''$). Moreover, in order to use only isolated stars, we excluded all sources with a secondary star within $\sim 0.5''$ in the GALACTICNUCLEUS catalogue (SIRIUS is a seeing-limited survey with an angular resolution $\sim 1''$). We excluded all stars with an uncertainty $>$ 0.05 mag in the GALACTICNUCLEUS catalogue or $> 0.1$ mag in the  SIRIUS catalogue. The calculation of the ZP was done using a two-sigma clipping algorithm to remove outliers. Figure\,\ref{ZP_second} shows all the common stars and those selected for calibration.

\subsubsection{Zero-point uncertainties}

\begin{figure}
   \includegraphics[width=\linewidth]{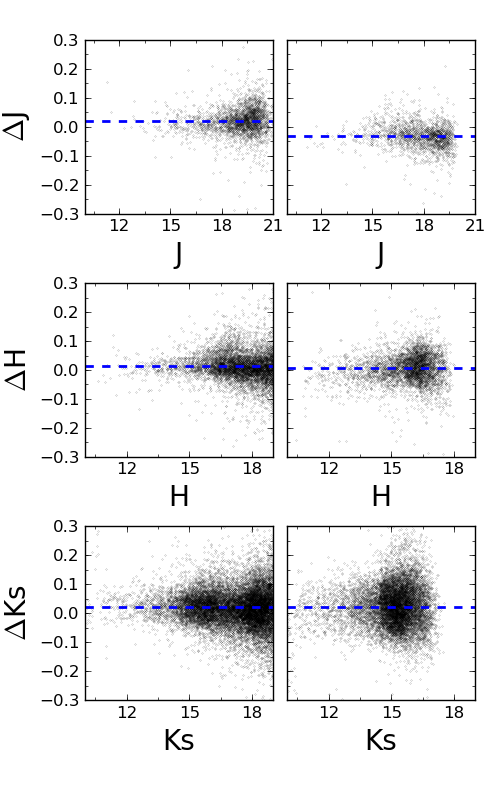}
   \caption{Left panels: Photometric comparison of the common stars between the central and the NSD East catalogues. Right panels: Photometric comparison Central-NSD West catalogues. The uncertainties were calculated using Eq. \ref{eq}. The blue dashed line indicates the mean offset existing between the considered regions.}
   
   \label{overlap_central_sides}
    \end{figure}

To estimate the uncertainties of the ZPs, we took several effects into account: 1) Firstly, the ZP uncertainty associated to the SIRIUS catalogue,  0.03 mag in all three bands \citep{Nishiyama:2006ai,Nishiyama:2008qa}, must be taken into account. 2) Also, the process used to combine the chips and pointings until obtaining the final catalogue is complex and may lead to additional systematic uncertainties of the ZP. 
 Figure\,\ref{overlap_central_sides} shows the comparison between the common stars in all three bands between the central and the NSD East and West catalogues. The uncertainties were computed using Eq. \ref{eq}. From this comparison we estimated an upper limit for the differential ZP uncertainty of $\sim 0.03$ mag in all three bands using a two-sigma clipping algorithm. This is consistent with the uncertainty estimated in \citet{Nogueras-Lara:2018aa} for the central field obtained comparing the observations taken for the central field of the survey (F1) and observations of the same region from a pilot study using also the speckle holography technique. 
3) The statistical uncertainty of the ZPs must also be calculated, but the high number of stars used for the ZP calibration at each band means that this uncertainty is negligible.

Combining quadratically the uncertainties (1) and (2) we ended up with an uncertainty of the ZP of $0.04$\,mag in all three bands, which is in agreement with the value of 0.036 mag that was obtained for the central field in \citet{Nogueras-Lara:2018aa}.

\subsection{Completeness}

Given the high number of sources detected, using the standard approach of inserting artificial stars would  enormously increase the computational time needed to analyse the regions covered. Instead of this, we used an alternative approach based on the determination of the critical distance from a bright star at which we are able to detect a star of any given magnitude \citep{Eisenhauer:1998tg,Harayama:2008ph}. This information is later translated to completeness maps for each magnitude bin considered. This technique assumes that the detection probability of a single star is constant across the analysed field. Since our final catalogues are produced combining independent pointings that were obtained under non-uniform observing conditions and since the stellar density varies as a function of position in the GC,  we analysed  the completeness due to crowding on smaller subregions of $2' \times 1.4'$. For a rough assessment of overall completeness, we subsequently averaged over all the subregions. We applied this technique only to the central pointings (1-15) as they are the most challenging ones in terms of crowding. Crowding is the factor that dominates completeness in the sensitive $K_{s}$- and $H$-band images. We estimated a  completeness of $\sim 80$ \% at $K_s \sim 16$ mag and $H \sim 18$ mag. Crowding is of little significance in the  $J$-band images because the much higher interstellar extinction (we note that the reddening induced colours  $J-K_{s}\approx5$ at the GC) at short NIR wavelengths reduces the number of detected stars considerably.
We ran artificial stars tests in the central pointing of the central catalogue (F1), the most crowded one,  and obtained a completeness of $\sim 80$ \% at $J \sim 20$ mag. This is a conservative limit for completeness due to crowding because most other fields will be less crowded.

We also ran the same simulations for $H$ and $K_s$ bands and confirmed the limits previously obtained. Moreover, a comparison with the deeper photometry obtained for the nuclear star cluster in $H$ and $K_s$ ($\sim 0.06''$ angular resolution) with NACO/VLT \citep{Gallego-Cano:2018aa} agrees with the completeness limits that we estimated here.

\subsection{Final catalogues}

We obtained accurate photometry in the NIR $J$, $H,$ and $K_s$ bands for $\sim 3.3 \times 10^6$ stars. Approximately 20\% of them where detected in $J$, 65\% in $H,$ and 90\% in $K_s$. The catalogue covers a total area of $\sim0.3$ square degrees, which corresponds to $\sim 6000$ pc$^2$. This supposes an average stellar density of $\sim 600$ stars/pc$^2$. 

The final catalogues published in this paper include 25 columns that contain information about the following aspects.

\begin{itemize}

\item {\bf Position and uncertainties between bands}: Right ascension and declination expressed in degrees (columns 1 and 3) and their corresponding uncertainties expressed in arcseconds (columns 2 and 4). For stars detected in more than one band these coordinates have been calculated averaging the positions of the detections. The uncertainties refer to the deviation of the measurements (see column 4 in Fig.\,\ref{pos_uncer}). For stars detected in a single band, we indicated the corresponding coordinates and the uncertainty associated to the detection (see column 3 in Fig.\,\ref{pos_uncer}).

\item {\bf Relative position and uncertainties for each band}: We included the positions and the associated uncertainties estimated from multiple detections of the same star in different pointings within the same band (column 3 in Fig.\,\ref{pos_uncer}). The positions are expressed in degrees and the uncertainties in units of arcseconds (columns 5-8 for $J$, 9-12 for $H$, 13-16 for $K_{s}$). A value of zero means that a star was not detected in a given band or that it was not detected multiple times in overlapping pointings. In those cases we refer to the previously derived upper limit of 30 milliarcsec (Sect. \ref{astrometry}).\\

\item {\bf Photometry and uncertainty for each band}: For each star we included the photometry in $J$, $H$, and $K_s$ and the associated uncertainties expressed in mag  (columns 17-22). A value of 99 indicates a non-detection.\\

\item {\bf Number of multiple detections}: This value indicates the number of multiple detections in overlapping pointings for each band (columns 23, 24, 25).\\

\end{itemize}

Table \ref{final_cat} outlines the first rows of the central catalogue specifying the columns described previously.

\setlength{\tabcolsep}{3.5pt}
\def\arraystretch{1.5}
       \begin{sidewaystable*}
       \footnotesize
       
\begin{center}
\caption{Fields included in each catalogue.}
\label{final_cat} 
 \begin{tabular}{ccccccccccccccccccccccccc}
 &  &  &  &  &  &  &  &  &  &  &  &  &  &  &  &  &  &  &  &  &  &  &  & \tabularnewline
\hline 
\hline 
ra & $\Delta$ra & dec & $\Delta$dec & ra$_J$ & $\Delta$ra$_J$ & dec$_J$ & $\Delta$dec$_J$ & ra$_H$ & $\Delta$ra$_H$ & dec$_H$ & $\Delta$dec$_H$ & ra$_H$ & $\Delta$ra$_K{_s}$ & dec$_K{_s}$  & $\Delta$dec$_K{_s}$  & $J$ & $dJ$ & $H$ & $dH$ & $K_s$ & $dK_s$ & $i_J$ & $i_H$ & $i_K{_s}$\tabularnewline
 $^\circ$& $''$ & $^\circ$  &  $''$  & $^\circ$ &  $''$  &  $^\circ$&  & $^\circ$ &  $''$ &$^\circ$  &  $''$  & $^\circ$ &  $''$  & $^\circ$ &  $''$  & mag & mag & mag & mag & mag & mag &  &  & \tabularnewline
\hline 
266.47 & 0.003 & -28.97 & 0.003 & 266.47 & 0.007 & -28.97 & 0.010 & 266.47 & 0.003 & -28.97 & 0.004 & 266.48 & 0.004 & -28.97 & 0.006 & 11.355 & 0.008 & 10.666 & 0.008 & 10.409 & 0.006 & 4 & 3 & 3\tabularnewline
266.47 & 0.002 & -28.97 & 0.002 & 266.47 & 0.010 & -28.97 & 0.012 & 266.47 & 0.007 & -28.97 & 0.010 & 266.48 & 0.005 & -28.97 & 0.006 & 12.148 & 0.009 & 11.721 & 0.007 & 11.556 & 0.007 & 4 & 4 & 4\tabularnewline
266.47 & 0.001 & -28.98 & 0.001 & 266.47 & 0.006 & -28.98 & 0.007 & 266.47 & 0.005 & -28.98 & 0.006 & 266.47 & 0.007 & -28.98 & 0.008 & 13.813 & 0.008 & 12.697 & 0.007 & 12.340 & 0.011 & 4 & 4 & 4\tabularnewline
266.48 & 0.006 & -28.96 & 0.007 & 266.48 & 0.002 & -28.96 & 0.003 & 266.48 & 0.006 & -28.96 & 0.007 & 266.48 & 0.003 & -28.96 & 0.004 & 14.063 & 0.006 & 13.524 & 0.015 & 13.411 & 0.008 & 3 & 4 & 3\tabularnewline
266.47 & 0.003 & -28.97 & 0.004 & 266.47 & 0.002 & -28.97 & 0.002 & 266.47 & 0.008 & -28.97 & 0.010 & 266.47 & 0.007 & -28.97 & 0.008 & 14.612 & 0.011 & 13.848 & 0.014 & 13.621 & 0.006 & 4 & 4 & 4\tabularnewline
266.48 & 0.001 & -28.97 & 0.002 & 266.48 & 0.005 & -28.97 & 0.006 & 266.48 & 0.000 & -28.97 & 0.000 & 266.48 & 0.007 & -28.97 & 0.008 & 14.753 & 0.009 & 12.861 & 0.006 & 12.051 & 0.008 & 4 & 4 & 4\tabularnewline
266.48 & 0.001 & -28.97 & 0.001 & 266.48 & 0.005 & -28.97 & 0.006 & 266.48 & 0.007 & -28.97 & 0.009 & 266.48 & 0.003 & -28.97 & 0.004 & 15.037 & 0.008 & 12.949 & 0.007 & 11.968 & 0.005 & 4 & 4 & 4\tabularnewline
266.48 & 0.007 & -28.97 & 0.009 & 266.48 & 0.006 & -28.97 & 0.007 & 266.48 & 0.008 & -28.97 & 0.010 & 266.48 & 0.027 & -28.97 & 0.028 & 15.146 & 0.009 & 14.287 & 0.011 & 13.961 & 0.032 & 4 & 4 & 4\tabularnewline
266.47 & 0.008 & -28.97 & 0.007 & 266.47 & 0.007 & -28.97 & 0.008 & 266.47 & 0.001 & -28.98 & 0.002 & 266.47 & 0.024 & -28.97 & 0.023 & 15.139 & 0.024 & 11.580 & 0.010 & 10.167 & 0.015 & 3 & 2 & 2\tabularnewline
266.47 & 0.007 & -28.96 & 0.008 & 266.47 & 0.004 & -28.96 & 0.004 & 266.47 & 0.003 & -28.96 & 0.003 & 266.47 & 0.024 & -28.96 & 0.025 & 15.328 & 0.012 & 11.605 & 0.010 & 10.029 & 0.021 & 3 & 3 & 2\tabularnewline
266.48 & 0.003 & -28.97 & 0.003 & 266.48 & 0.007 & -28.97 & 0.008 & 266.48 & 0.003 & -28.97 & 0.004 & 266.48 & 0.006 & -28.97 & 0.007 & 15.389 & 0.008 & 14.787 & 0.008 & 14.575 & 0.011 & 4 & 4 & 4\tabularnewline
266.48 & 0.002 & -28.97 & 0.002 & 266.48 & 0.006 & -28.97 & 0.007 & 266.48 & 0.006 & -28.97 & 0.007 & 266.48 & 0.002 & -28.97 & 0.002 & 15.405 & 0.007 & 13.721 & 0.007 & 12.859 & 0.006 & 4 & 4 & 4\tabularnewline
266.47 & 0.001 & -28.97 & 0.001 & 266.47 & 0.007 & -28.97 & 0.009 & 266.47 & 0.006 & -28.97 & 0.007 & 266.47 & 0.003 & -28.97 & 0.004 & 15.424 & 0.008 & 14.649 & 0.007 & 14.400 & 0.005 & 4 & 4 & 4\tabularnewline
266.47 & 0.007 & -28.96 & 0.007 & 266.47 & 0.020 & -28.96 & 0.020 & 266.47 & 0.020 & -28.96 & 0.015 & 266.47 & 0.005 & -28.96 & 0.006 & 15.735 & 0.010 & 12.232 & 0.008 & 10.437 & 0.014 & 4 & 4 & 4\tabularnewline
266.48 & 0.001 & -28.96 & 0.002 & 266.48 & 0.003 & -28.96 & 0.004 & 266.48 & 0.006 & -28.96 & 0.007 & 266.48 & 0.002 & -28.96 & 0.002 & 15.762 & 0.011 & 14.748 & 0.007 & 14.362 & 0.024 & 4 & 4 & 4\tabularnewline
266.48 & 0.007 & -28.97 & 0.007 & 266.48 & 0.002 & -28.97 & 0.002 & 266.48 & 0.017 & -28.98 & 0.019 & 266.48 & 0.017 & -28.97 & 0.019 & 15.851 & 0.008 & 14.989 & 0.007 & 14.701 & 0.010 & 4 & 4 & 4\tabularnewline
266.48 & 0.001 & -28.97 & 0.002 & 266.48 & 0.007 & -28.97 & 0.008 & 266.48 & 0.009 & -28.97 & 0.011 & 266.48 & 0.006 & -28.97 & 0.007 & 15.946 & 0.008 & 15.030 & 0.007 & 14.695 & 0.005 & 4 & 4 & 4\tabularnewline
266.48 & 0.001 & -28.97 & 0.002 & 266.48 & 0.005 & -28.97 & 0.006 & 266.48 & 0.003 & -28.97 & 0.004 & 266.48 & 0.004 & -28.97 & 0.005 & 15.960 & 0.008 & 13.629 & 0.007 & 12.568 & 0.009 & 4 & 4 & 4\tabularnewline
266.48 & 0.003 & -28.97 & 0.003 & 266.48 & 0.001 & -28.97 & 0.001 & 266.48 & 0.007 & -28.97 & 0.008 & 266.48 & 0.008 & -28.97 & 0.010 & 16.069 & 0.009 & 15.204 & 0.011 & 14.895 & 0.009 & 4 & 3 & 3\tabularnewline
266.48 & 0.008 & -28.97 & 0.011 & 266.48 & 0.003 & -28.97 & 0.004 & 266.48 & 0.024 & -28.97 & 0.027 & 266.48 & 0.027 & -28.97 & 0.017 & 16.538 & 0.038 & 15.649 & 0.058 & 15.854 & 0.038 & 3 & 2 & 3\tabularnewline
266.47 & 0.001 & -28.97 & 0.001 & 266.47 & 0.003 & -28.98 & 0.004 & 266.47 & 0.002 & -28.98 & 0.003 & 266.47 & 0.003 & -28.98 & 0.004 & 16.371 & 0.008 & 15.469 & 0.007 & 15.267 & 0.030 & 4 & 3 & 4\tabularnewline
266.48 & 0.004 & -28.97 & 0.005 & 266.48 & 0.001 & -28.97 & 0.001 & 266.48 & 0.008 & -28.97 & 0.010 & 266.48 & 0.005 & -28.97 & 0.006 & 16.310 & 0.008 & 12.450 & 0.007 & 10.500 & 0.010 & 4 & 3 & 4\tabularnewline
266.47 & 0.005 & -28.97 & 0.006 & 266.47 & 0.005 & -28.97 & 0.006 & 266.47 & 0.007 & -28.97 & 0.009 & 266.47 & 0.002 & -28.97 & 0.003 & 16.395 & 0.017 & 15.688 & 0.022 & 15.505 & 0.010 & 4 & 4 & 3\tabularnewline
... & ... & ... & ... & ... & ... & ... & ... & ... & ... & ... & ... & ... & ... & ... & ... & ... & ... & ... & ... & ... & ... & ... & ... & ...\tabularnewline
\hline 
\end{tabular}
\end{center}
\vspace{0.4cm}

\textbf{Notes.} First rows of the central catalogue. ra, $\Delta$ra  and dec, $\Delta$dec are the right ascension and the declination of the sources (and their corresponding uncertainties) obtained combining the detections in several bands.  ra$_i$, $\Delta$ra$_i$  and dec$_i$, $\Delta$dec$_i$ correspond to the coordinates and uncertainties for any single band, $i$. The coordinates were rounded to two decimals in this table. $J$, $dJ$, $H$, $dH$, $K_s$ and $dK_s$ are the photometry and the associated uncertainties. A 99 indicates that there is no detection for a given source.  $i_J$, $i_H$ and $i_{K_s}$ indicate the number of fields where a star was detected to obtain its final photometry.

\end{sidewaystable*}

\begin{sidewaysfigure*}
   \includegraphics[width=\linewidth]{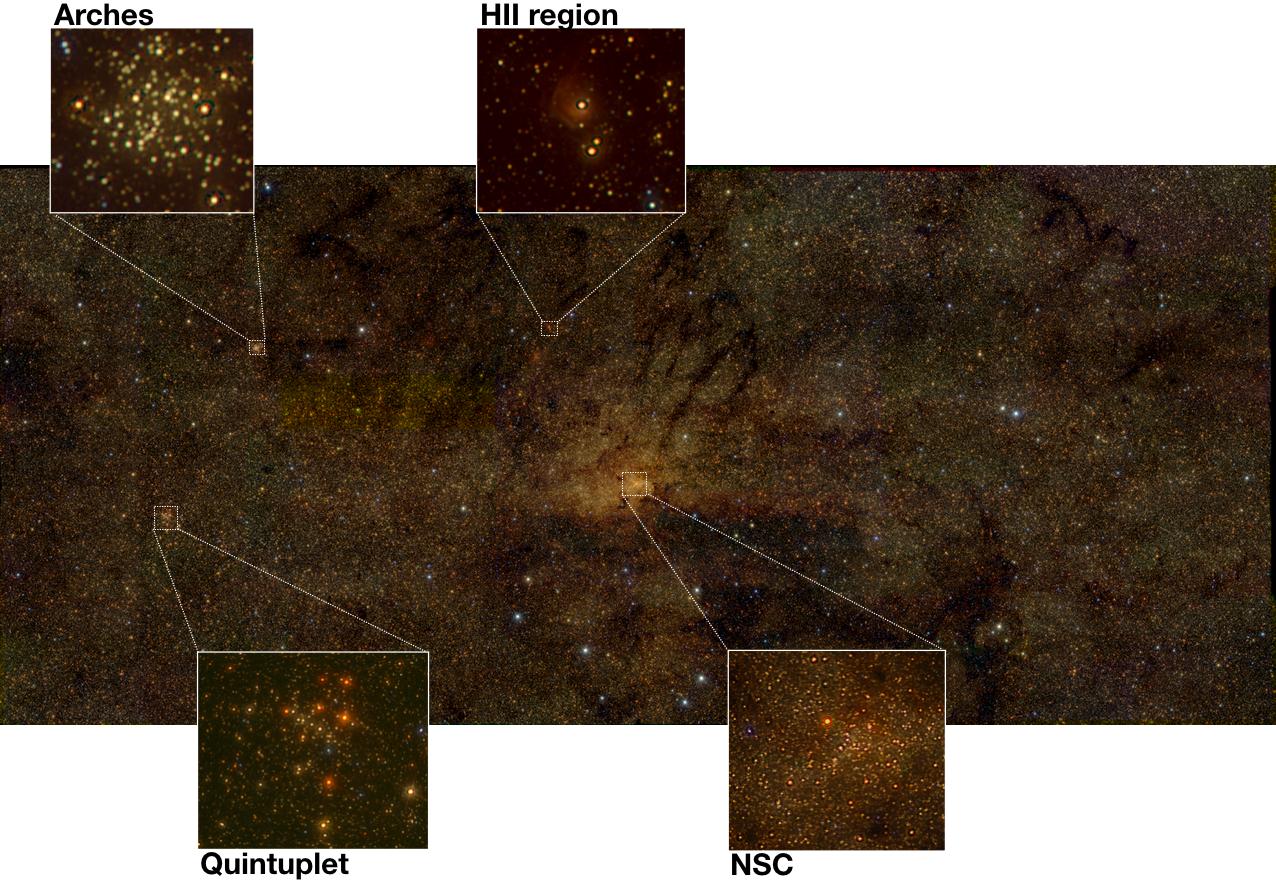}
   \caption{RGB image (red = $K_s$ band, green = $H$ band and blue = $J$ band) of the central catalogue corresponding to fields from 1 to 30. Field 7 is not included in $J$ band since the observing conditions were not acceptable to include it in the survey. The insets show the NSC \citep[e.g.][]{Launhardt:2002nx,Schodel:2014fk,Boehle:2016aa,Gravity-Collaboration:2018aa} , the Arches cluster \citep[e.g.][]{Martins:2008aa,Clarkson:2012fk}, the Quintuplet cluster \citep[e.g.][]{Figer:1999uq,Najarro:2009vn}, and one of the HII regions that we observed. The size of the field is $\sim 35' \times 20'$ and the orientation is galactic north up, galactic east to the left.}
 
   \label{mosaic}
    \end{sidewaysfigure*}

Figure \ref{mosaic} shows an RGB image corresponding to the GC Central catalogue produced from a mosaic of pointings 1 to 30.

    The catalogues described in this paper are made publicly available at the CDS via anonymous ftp to cdsarc.u-strasbg.fr or via http://cdsarc.u-strasbg.fr/. Moreover, we are now preparing the individual catalogues for each chip and band for the ESO phase-three data release of our ESO Large Programme that underlies this catalogue. The data will be available from the ESO Science Archive.

\section{Quality assessment}

We carried out several tests to check the photometric quality of our catalogues. These tests are described below.

\subsection{Comparison with SIRIUS catalogue}

\begin{figure}
   \includegraphics[width=\linewidth]{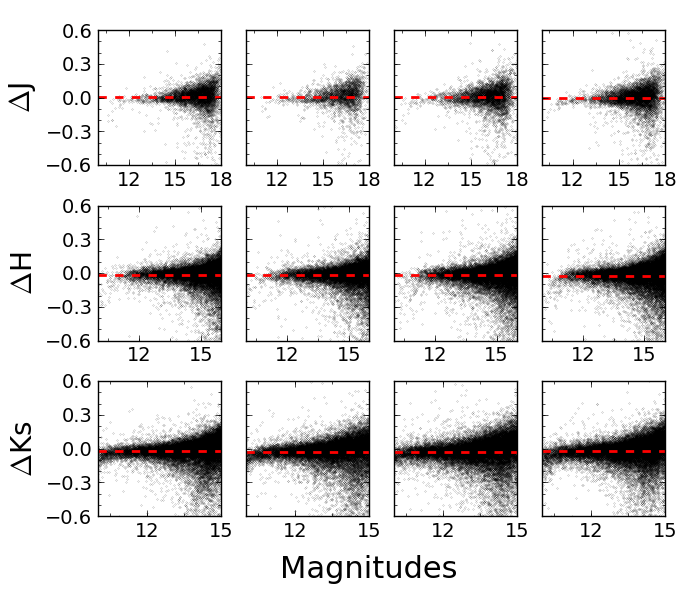}
   \caption{Photometric comparison between GALACTICNUCLEUS and the SIRIUS IRSF survey across the central catalogue. We divided the central catalogue into four equal regions from east to west along the Galactic Plane; these are shown in each of the columns for all three bands. The red dashed line indicates the photometric offset.}
   
   \label{variation_position}
    \end{figure}

 We compared the photometry obtained for all three bands with the SIRIUS IRSF survey catalogue to check that the ZP does not change significantly across the final region covered by the catalogues when combining the different fields. For this purpose we used the GC Central catalogue, which we consider to make a good test case because it contains the most complex region and combines 30 different pointings observed under different seeing conditions at different epochs (see Table\,\ref{obs_cen}).  We cross-matched both catalogues and identified common stars in all three bands. We divided the GC central region into four independent vertical columns, from east to west along the Galactic Plane,  and analysed them independently. We used Eq.\,\ref{eq} to compare the photometry. Figure \ref{variation_position} shows the results. We calculated the offset using a two-sigma clipping algorithm to remove outliers and avoid saturated and overly faint stars. We considered stars of $J=12.5 - 15.5$ mag, $H= 12.3 -  15.5$ mag, and $K_{s}=12.5 - 14$ mag (see also Fig.\,\ref{ZP_second}). We then computed the difference between the maximum and minimum offsets for each band between the four different columns. This difference is $<0.01$ mag in all three bands. We conclude that there is no significant variability of the ZP across the GC Central field.\\

\subsection{Comparison with a previously used sub-catalogue}

\begin{figure}
   \includegraphics[width=\linewidth]{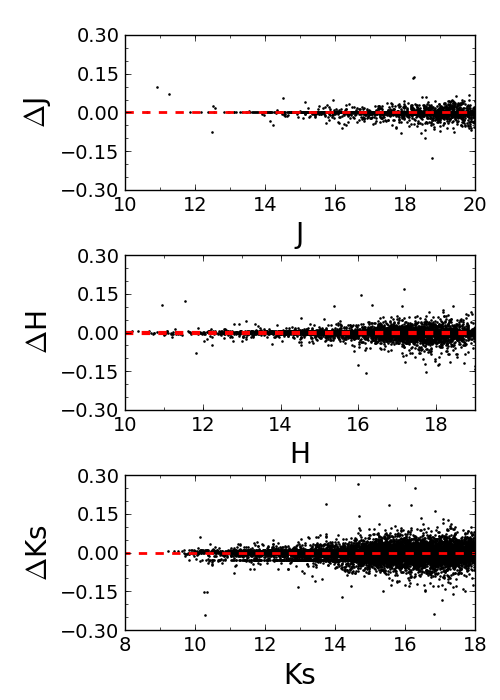}
   \caption{Photometric comparison between the final version of the central catalogue of GALACTICNUCLEUS and a preliminary version obtained using the fields from 1-15 (except field 7). The red dashed line indicates the offset between the photometric ZPs.}
   
   \label{comparison_15_30}
    \end{figure}

We compared the final GC Central catalogue with a previous version that included only 14 of 30  fields and that was used in previous work (Nogueras-Lara et al., submitted; Gallego-Cano et al., submitted). The main difference introduced by the final version is the photometry of the common stars detected in pointings  that were not included in the former version. Also, the ZPs computed at intermediate steps could have varied (see Sect. \ref{rec_zp}) because less fields were present in the previous version of the GC Central catalogue. Figure\,\ref{comparison_15_30} shows the comparison between the preliminary and the final catalogues following Eq.\,\ref{eq}. Both catalogues agree very well as expected.\\

\subsection{Comparison with VVV}

\begin{figure}
   \includegraphics[width=\linewidth]{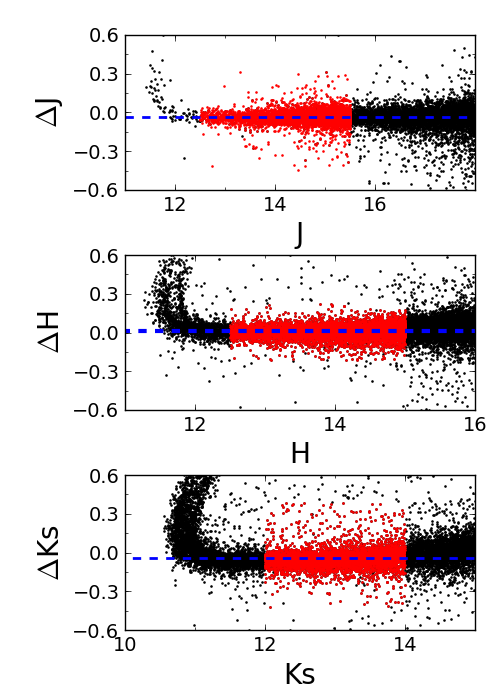}
   \caption{Photometric comparison between the final version of the central catalogue of GALACTICNUCLEUS and the VVV catalogue with aperture photometry. The red dots indicate the stars used to compute the offset. The blue dashed line corresponds to the photometric offset between both catalogues. The bright tails are due to the strong saturation in the VVV catalogue.}
   
   \label{comparison_vvv}
    \end{figure}

Up to now, the most complete multi-band catalogue to study the GC was the VVV survey \citep{Minniti:2010fk,Saito:2012fk}. This survey covers an area of 520 deg$^2$ in the Galactic Bulge and a section of the mid-plane with high star formation activity. It uses aperture photometry to characterise $\sim 10^9$ sources in the NIR. Since the $JHK_s$ filters are very close to those of GALACTICNUCLEUS, we compared the photometry of common stars to analyse the systematic errors of the ZP and the data quality of our survey. We used the VVV aperture photometry computed in a radius of 1$''$. To calculate the ZP offset, we removed all the stars with an uncertainty $>0.05$ mag in both catalogues and in all three bands. Due to the extreme source crowding in the $H$ and $K_s$ bands, we excluded all stars with a fainter counterpart detected in GALACTICNUCLEUS within a radius of 1$''$. We used a two-sigma clipping algorithm to remove outliers and avoid saturated and overly faint stars as indicated by the red dots in Fig. \ref{comparison_vvv}. We found that the small shifts agree with the systematic uncertainties estimated for the ZP. Moreover, we verified that a comparison between SIRIUS and VVV gives a similar result.

\begin{figure}
   \includegraphics[width=\linewidth]{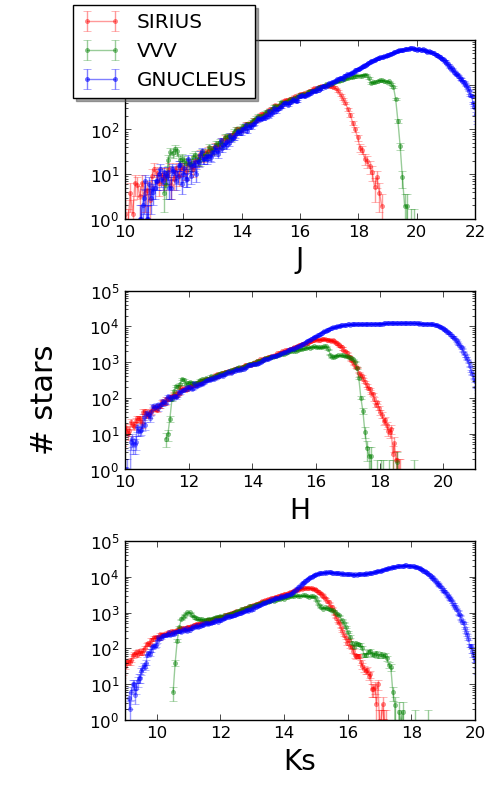}
   \caption{Luminosity functions obtained with the SIRIUS survey (in red), the VVV survey (in green), and the GALACTICNUCLEUS survey (in blue). The uncertainties are Poisson errors (square root of the number of stars in each bin).}
   
   \label{VVV_LFs}
    \end{figure}

Concomitantly, we compared the $J$, $H,$ and $K_s$ luminosity functions (LFs) from VVV, SIRIUS and GALACTICNUCLEUS obtained for the central catalogue (Fig.\,\ref{VVV_LFs}). GALACTICNUCLEUS  reaches $\gtrsim1$ mag deeper in $J$ and $\gtrsim2$\,mag in $H$ and $K_s$. The GALACTICNUCLEUS survey therefore significantly improves the number of stars detected at the faint end of the $H$ and $K_s$ LFs. The GALACTICNUCLEUS LFs are also more complete at the bright end, where the VVV survey suffers from far stronger saturation (green bumps at the bright end in Fig. \ref{comparison_vvv}) because of the larger pixel scale of VIRCAM  and the longer exposure time used in VVV. In particular, the GALACTICNUCLEUS catalogue completely covers the RC bump in $K_s$. Red clump stars are important standard candles and tracers of Galactic structure. Multiple RC bumps can give information on recent star formation events or on different overlapping structures along the LoS \citep[e.g.][]{Girardi:2016fk}. The SIRIUS survey can also be seen to be a better reference for photometric calibration in comparison to the VVV catalogue since it is far less saturated and covers a larger magnitude range in $H$ and $K_s$ bands.

Finally, we compared a region in the central catalogue (centred on 17$^h$ 45$^m$ 47.524$^s$, -28$^\circ$ 55$'$ 11.847$''$ and covering an area of $\sim 20'\times 17'$) with the recent update of the VVV survey using PSF photometry \citep{Alonso-Garcia:2018aa}. We obtained that the VVV photometry is relatively incomplete in the bright part and does not improve the faint end of the LFs obtained using the aperture photometry. This is mainly because the new release of the catalogue only accepts a star if it is detected in at least three bands (out of $Z,Y,J,H,K_s$). Moreover, this survey has not been specially designed for the GC. For the same region we also compared the photometric uncertainties given by the VVV survey with PSF photometry and GALACTICNUCLEUS. We analysed both catalogues in the same magnitude range to avoid a bias due to the different completeness limits. Therefore, we considered only stars with $J < 19$ mag, $H < 16$ mag, and $K_s < 14$ mag. We obtained a mean uncertainty of $\Delta_J = 0.06$ mag, $\Delta_H = 0.02$ mag, and $\Delta_K{_s} = 0.02$ mag for  VVV and  $\Delta_J = 0.02$ mag, $\Delta_H = 0.01$ mag, and $\Delta_K{_s} = 0.01$ mag for GALACTICNUCLEUS. This represents an improvement by factors of between two and three.

\subsection{Comparison with NICMOS HST}

\begin{figure*}
       \begin{center}
   \includegraphics[scale=0.4]{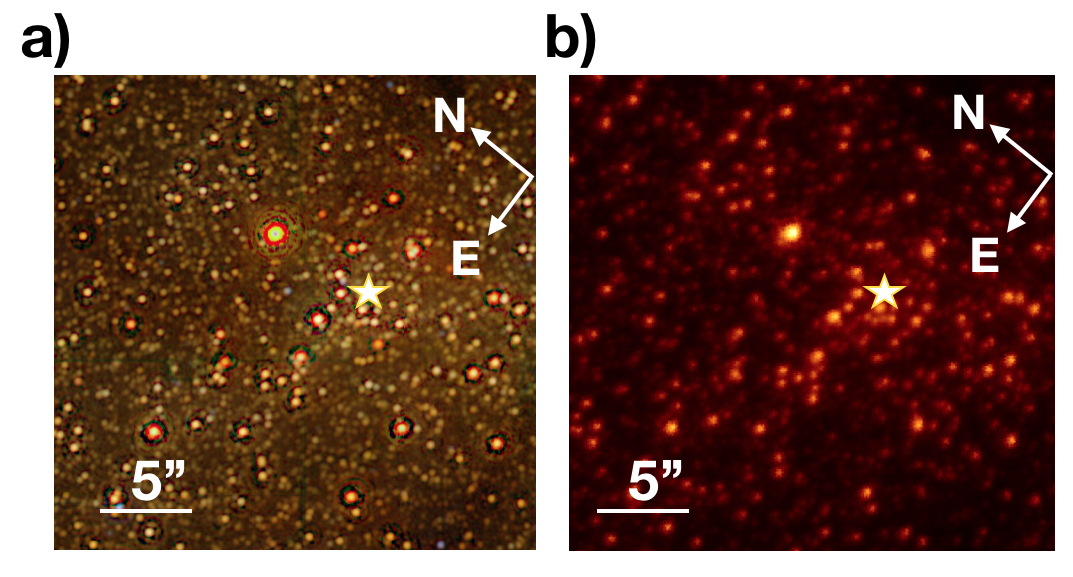}
   \caption{Comparison of the NSC between (a) GALACTICNUCLEUS ($JHK_s$ RGB image) and NICMOS HST F190 (b). The white star indicates the position of Sgr A*.}
   
   \label{NICMOS}
      \end{center}
    \end{figure*}

We compared our data with the NICMOS HST survey carried out by \citet{Dong:2011ff} using the narrow band filters F187 and F190. We selected a common region between both catalogues of $\sim 20'\times 10'$ centred on Sgr A*. We detected around three times more sources in the GALACTICNUCLEUS survey. We also compared the uncertainties and obtained that the relative uncertainty of the NICMOS HST survey is $\sim 6\%$ for F187 and $\sim 5\%$ for F190, whereas we reached a relative median uncertainty of $\sim 2 \%$ in all three bands for GALACTICNUCLEUS. Moreover, we are able to achieve the same angular resolution of $\sim 0.2''$ using a ground-based telescope. Figure \ref{NICMOS} shows the comparison between the NSC as it is seen by NICMOS and by GALACTICNUCLEUS.

\subsection{Detailed analysis of pointing F1}
Additional tests were carried out for the most crowded field (pointing F1) in the first paper of this series \citep{Nogueras-Lara:2018aa}. We checked the photometry and the photometric uncertainties comparing the field with previous data from 2013. We also used calibration stars to check the variation of the ZP across each detector, and estimated the uncertainties using simulations to test the influence of the extreme crowding on the photometric errors. In general, all the conclusions obtained for pointing F1 can be used as upper limits for the rest of the fields under the same observing conditions.

\section{Limitations of the survey}

As was shown previously, the GALACTICNUCLEUS catalogue represents a great improvement in the state of art available to study the stellar structure and population of the nuclear bulge of the Milky Way. In spite of this, there are two main limitations that need to be taken into account. 

\begin{itemize}

\item {\bf Photometric saturation}: In spite of the short exposure times used for the observations, the very good seeing conditions (see Tables \ref{obs_cen}, \ref{obs_bul} and \ref{obs_disc}) provoke a significant saturation of the bright sources in $K_s$ band. To study this effect, we created $K_s$ LFs using GALACTICNUCLEUS and SIRIUS data. Because of the smaller size of the telescope used for SIRIUS (1.4 m), the effect of saturation is less  important in comparison to GALACTICNUCLEUS. This comparison shows that for $K_s \sim 11.5$ mag the GALACTICNUCLEUS photometry is not reliable (see Fig.\,\ref{ZP_second}). Saturation is less important at $H$ and almost negligible at $J$. \\

\item {\bf Restrictive criterion for accepting stars}: In order to estimate the uncertainties for individual stars and to avoid spurious detections, we imposed that only stars detected in all three sub-images (see section\,\ref{summary}) are accepted \citep[see also][]{Nogueras-Lara:2018aa}. This can produce an estimated loss of $\sim 30-40$ \% of the total number of stars present in the final holographic products. We are working on a bootstrapping algorithm to estimate the uncertainties and to obtain a deeper photometry improving the threshold for the detection of faint stars, to be published later.

\end{itemize}

\section{Colour-magnitude diagrams}

\begin{figure*}
   \includegraphics[width=\linewidth]{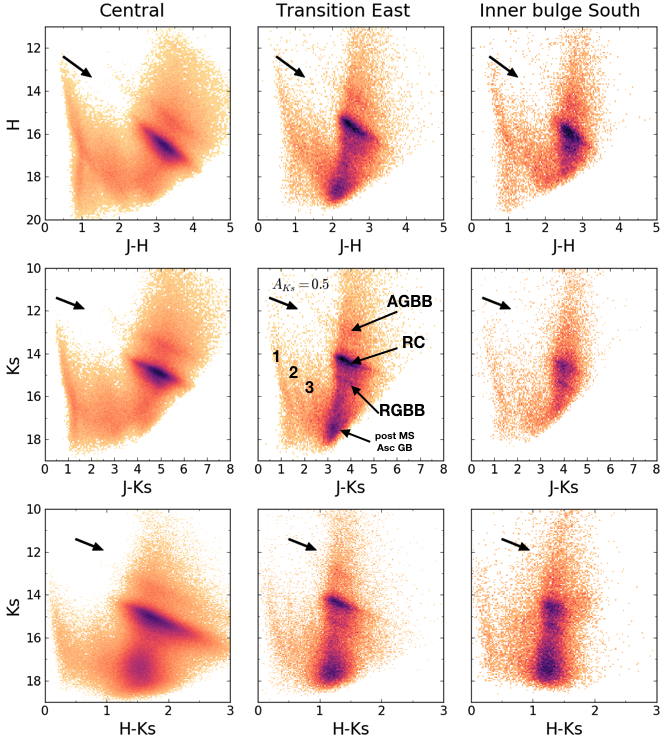}
   \caption{Colour-magnitude diagrams for different regions of the GALACTICNUCLEUS survey. The colour code corresponds to stellar densities, using a power stretch scale. First, second, and third columns correspond to the central, the transition East, and the inner bulge South catalogues, respectively. The black arrow indicates the reddening vector and an extinction of 0.5 mag in the $K_s$ band. The labels in the central panel correspond to the different stellar populations present in the CMDs: AGBB (asymptotic giant branch bump), RC (red clump), RGBB (red giant branch bump), post MS (post main sequence stars), and Asc GB (ascending giant branch) and the numbers from 1 to 3 indicate the foreground population corresponding to three spiral arms.}
   
   \label{CMD}
    \end{figure*}

The GALACTICNUCLEUS survey covers regions in the inner bulge, the NB, and the transition region between them. The different stellar populations in these regions combined with the highly variable interstellar extinction result in significant differences between the corresponding  colour-magnitude diagrams (CMDs). Figure\,\ref{CMD} shows the CMDs obtained for the GC Central, the Transition East, and the Inner Bulge South catalogues (columns one, two and three, respectively). Typical types of stars found in the CMD are specified in the central panel: The foreground population ($J-H \lesssim 2$ mag, $J-K_s \lesssim 3$ mag and $H-K_s \lesssim 1$ mag) corresponds to stars along the LoS from Earth to the GC probably tracing three spiral arms \citep{Nogueras-Lara:2018aa}. The stellar population located at $J-H \sim 2-3$  mag, $J-K_s \sim 3-5$ mag, $H-K_s \sim 1-2$ mag, and $H \sim 15$ mag, and $K_s \sim 13$ mag is composed of stars in the asymptotic giant branch (AGB) bump. The prominent feature located at $J-H \sim 2-3$ mag, $J-K_s \sim 3-5$ mag, $H-K_s \sim 1-2$ mag and $H \sim 16$ mag and $K_s \sim 15$ mag corresponds to the RC \citep [giant stars in their helium-core-burning sequence][]{Girardi:2016fk} and the red giant branch bump \citep[RGBB, see, e.g.][]{Cassisi:1997aa,Salaris:2002ab,Nataf:2014aa,Nogueras-Lara:2018ab}. Previous studies analysing regions from the Galactic bulge at higher latitudes ($\sim 1^\circ$) also found a secondary RC feature and identified it as a stellar population tracing a spiral arm beyond the GC \citep[][]{Gonzalez:2011ac,Gonzalez:2018aa}. This feature coincides with what we here identify as the RGBB. Nevertheless, given the extreme stellar density and crowding in the low-latitude regions covered by the GALACTICNUCLEUS survey, it is very unlikely to observe any star beyond the GC. Moreover our previous work \citep{Nogueras-Lara:2018ab} shows that the observed feature is compatible with the RGBB. The RC and the RGBB are of great interest since their relative properties can allow us  to constrain the metallicity and age of the stellar population in the sample \citep[e.g.][]{Nogueras-Lara:2018ab}. Finally, stars located at $K_s \gtrsim 17$ mag and below the RC and the RGBB belong to the ascending giant branch and post-main sequence.

\section{Conclusions}

We present the first public data release of the GALACTICNUCLEUS survey, a high-angular-resolution ($0.2''$) $JHK_s$ NIR imaging survey specially designed to study the GC. The surveyed area is $\sim0.3$ square degrees, which corresponds to $\sim 6000$ pc$^2$. We describe in detail the reduction and analysis process from the raw images to the final photometric catalogue. We obtained accurate photometry for more than 3 million stars superseding previous surveys for the GC and reaching $J \sim 1$ mag and $H, K_s \gtrsim 2$ mag deeper than the best current catalogues for the same region. We reached 5\,$\sigma$ detection limits of $J \sim 22$ mag, $H \sim 21$ mag, and $K_s \sim 21$ mag. The uncertainty of the photometric ZP is $\lesssim0.04$\,mag in all bands. Relative photometric uncertainties are $\lesssim 0.05$ mag at $J\lesssim21$ mag, $H\lesssim19$ mag, and $K_s\lesssim18$ mag. The absolute astrometric uncertainty is $\sim 0.05''$. All relative astrometric positions in the final catalogue are $<0.05''$. We note that the relative astrometric positions between stars on individual chips for any given pointing are more than an order of magnitude smaller, typically only $\sim$1\,milli-arcsecond for moderately bright stars. However, the uncertainties increase when combining the chips and pointings to the final catalogue.

We present what are probably the most complete  CMDs of the GC so far, covering a large area and different regions in the nuclear stellar disc and inner bulge. The CMDs show clear differences caused by differential extinction and distinct stellar populations. We identified the foreground population, the AGBB, the post main sequence, and the ascending giant branch.  The RC and the RGBB are also covered.

Sharing our catalogue with the community, we expect to widen its usage to solve a variety of problems and unsolved questions related to the stellar population and structure of the innermost part of the Milky Way.

  \begin{acknowledgements}
      The research leading to these results has received funding from
      the European Research Council under the European Union's Seventh
      Framework Programme (FP7/2007-2013) / ERC grant agreement
      n$^{\circ}$ [614922]. This work is based on observations made with ESO
      Telescopes at the La Silla Paranal Observatory under programmes
      IDs 195.B-0283 and 091.B-0418. We thank the staff of
      ESO for their great efforts and helpfulness. Author F N-L acknowledges financial support from the State
Agency for Research of the Spanish MCIU through the "Center of Excellence Severo
Ochoa" award for the Instituto de Astrof\'isica de Andaluc\'ia (SEV-2017-0709). F N-L acknowledges financial support from a MECD pre-doctoral contract, code FPU14/01700. F N acknowledges financial support through Spanish grants ESP2015-65597-C4-1-R and ESP2017-86582-C4-1-R (MINECO/FEDER). N.N. acknowledges support by Sonderforschungsbereich SFB 881 `The Milky Way System' (subproject B8) of the German Research Foundation (DFG).
\end{acknowledgements}

\bibliography{/Users/fnoguer/Documents/Doctorado/My_papers/BibGC.bib}

\begin{thebibliography}{44}
\expandafter\ifx\csname natexlab\endcsname\relax\def\natexlab#1{#1}\fi

\bibitem[{{Alonso-Garc{\'\i}a} {et~al.}(2018){Alonso-Garc{\'\i}a}, {Saito},
  {Hempel}, {Minniti}, {Pullen}, {Catelan}, {Ramos}, {Cross}, {Gonzalez},
  {Lucas}, {Palma}, {Valenti}, \& {Zoccali}}]{Alonso-Garcia:2018aa}
{Alonso-Garc{\'\i}a}, J., {Saito}, R.~K., {Hempel}, M., {et~al.} 2018, \aap,
  619, A4

\bibitem[{{Boehle} {et~al.}(2016){Boehle}, {Ghez}, {Sch{\"o}del}, {Meyer},
  {Yelda}, {Albers}, {Martinez}, {Becklin}, {Do}, \& {Lu}}]{Boehle:2016aa}
{Boehle}, A., {Ghez}, A.~M., {Sch{\"o}del}, R., {et~al.} 2016, \apj, 830, 17

\bibitem[{{Cassisi} \& {Salaris}(1997)}]{Cassisi:1997aa}
{Cassisi}, S. \& {Salaris}, M. 1997, \mnras, 285, 593

\bibitem[{{Clarkson} {et~al.}(2012){Clarkson}, {Ghez}, {Morris}, {Lu},
  {Stolte}, {McCrady}, {Do}, \& {Yelda}}]{Clarkson:2012fk}
{Clarkson}, W.~I., {Ghez}, A.~M., {Morris}, M.~R., {et~al.} 2012, \apj, 751,
  132

\bibitem[{{Crocker}(2012)}]{Crocker:2012fk}
{Crocker}, R.~M. 2012, \mnras, 423, 3512

\bibitem[{{Diolaiti} {et~al.}(2000){Diolaiti}, {Bendinelli}, {Bonaccini},
  {Close}, {Currie}, \& {Parmeggiani}}]{Diolaiti:2000qo}
{Diolaiti}, E., {Bendinelli}, O., {Bonaccini}, D., {et~al.} 2000, \aaps, 147,
  335

\bibitem[{{Dong} {et~al.}(2011){Dong}, {Wang}, {Cotera}, {Stolovy}, {Morris},
  {Mauerhan}, {Mills}, {Schneider}, {Calzetti}, \& {Lang}}]{Dong:2011ff}
{Dong}, H., {Wang}, Q.~D., {Cotera}, A., {et~al.} 2011, \mnras, 417, 114

\bibitem[{{Eisenhauer} {et~al.}(1998){Eisenhauer}, {Quirrenbach}, {Zinnecker},
  \& {Genzel}}]{Eisenhauer:1998tg}
{Eisenhauer}, F., {Quirrenbach}, A., {Zinnecker}, H., \& {Genzel}, R. 1998,
  \apj, 498, 278

\bibitem[{{Figer} {et~al.}(1999){Figer}, {McLean}, \& {Morris}}]{Figer:1999uq}
{Figer}, D.~F., {McLean}, I.~S., \& {Morris}, M. 1999, \apj, 514, 202

\bibitem[{{Fritz} {et~al.}(2011){Fritz}, {Gillessen}, {Dodds-Eden}, {Lutz},
  {Genzel}, {Raab}, {Ott}, {Pfuhl}, {Eisenhauer}, \&
  {Yusef-Zadeh}}]{Fritz:2011fk}
{Fritz}, T.~K., {Gillessen}, S., {Dodds-Eden}, K., {et~al.} 2011, \apj, 737, 73

\bibitem[{{Gallego-Cano} {et~al.}(2018){Gallego-Cano}, {Sch{\"o}del}, {Dong},
  {Nogueras-Lara}, {Gallego-Calvente}, {Amaro-Seoane}, \&
  {Baumgardt}}]{Gallego-Cano:2018aa}
{Gallego-Cano}, E., {Sch{\"o}del}, R., {Dong}, H., {et~al.} 2018, \aap, 609,
  A26

\bibitem[{{Girardi}(2016)}]{Girardi:2016fk}
{Girardi}, L. 2016, \araa, 54, 95

\bibitem[{{Gonzalez} {et~al.}(2018){Gonzalez}, {Minniti}, {Valenti},
  {Alonso-Garc{\'\i}a}, {Debattista}, {Zoccali}, {Rejkuba}, {Dias}, {Surot},
  {Hempel}, \& {Saito}}]{Gonzalez:2018aa}
{Gonzalez}, O.~A., {Minniti}, D., {Valenti}, E., {et~al.} 2018, \mnras, 481,
  L130

\bibitem[{{Gonzalez} {et~al.}(2011){Gonzalez}, {Rejkuba}, {Minniti}, {Zoccali},
  {Valenti}, \& {Saito}}]{Gonzalez:2011ac}
{Gonzalez}, O.~A., {Rejkuba}, M., {Minniti}, D., {et~al.} 2011, \aap, 534, L14

\bibitem[{{Gravity Collaboration} {et~al.}(2018){Gravity Collaboration},
  {Abuter}, {Amorim}, {Anugu}, {Baub{\"o}ck}, {Benisty}, {Berger}, {Blind},
  {Bonnet}, {Brandner}, {Buron}, {Collin}, {Chapron}, {Cl{\'e}net}, {Coud{\'e}
  Du Foresto}, {de Zeeuw}, {Deen}, {Delplancke-Str{\"o}bele}, {Dembet},
  {Dexter}, {Duvert}, {Eckart}, {Eisenhauer}, {Finger}, {F{\"o}rster
  Schreiber}, {F{\'e}dou}, {Garcia}, {Garcia Lopez}, {Gao}, {Gendron},
  {Genzel}, {Gillessen}, {Gordo}, {Habibi}, {Haubois}, {Haug}, {Hau{\ss}mann},
  {Henning}, {Hippler}, {Horrobin}, {Hubert}, {Hubin}, {Jimenez Rosales},
  {Jochum}, {Jocou}, {Kaufer}, {Kellner}, {Kendrew}, {Kervella}, {Kok},
  {Kulas}, {Lacour}, {Lapeyr{\`e}re}, {Lazareff}, {Le Bouquin}, {L{\'e}na},
  {Lippa}, {Lenzen}, {M{\'e}rand}, {M{\"u}ler}, {Neumann}, {Ott}, {Palanca},
  {Paumard}, {Pasquini}, {Perraut}, {Perrin}, {Pfuhl}, {Plewa}, {Rabien},
  {Ram{\'{\i}}rez}, {Ramos}, {Rau}, {Rodr{\'{\i}}guez-Coira}, {Rohloff},
  {Rousset}, {Sanchez-Bermudez}, {Scheithauer}, {Sch{\"o}ller}, {Schuler},
  {Spyromilio}, {Straub}, {Straubmeier}, {Sturm}, {Tacconi}, {Tristram},
  {Vincent}, {von Fellenberg}, {Wank}, {Waisberg}, {Widmann}, {Wieprecht},
  {Wiest}, {Wiezorrek}, {Woillez}, {Yazici}, {Ziegler}, \&
  {Zins}}]{Gravity-Collaboration:2018aa}
{Gravity Collaboration}, {Abuter}, R., {Amorim}, A., {et~al.} 2018, \aap, 615,
  L15

\bibitem[{{Harayama} {et~al.}(2008){Harayama}, {Eisenhauer}, \&
  {Martins}}]{Harayama:2008ph}
{Harayama}, Y., {Eisenhauer}, F., \& {Martins}, F. 2008, \apj, 675, 1319

\bibitem[{{Kissler-Patig} {et~al.}(2008){Kissler-Patig}, {Pirard}, {Casali},
  {Moorwood}, {Ageorges}, {Alves de Oliveira}, {Baksai}, {Bedin}, {Bendek},
  {Biereichel}, {Delabre}, {Dorn}, {Esteves}, {Finger}, {Gojak}, {Huster},
  {Jung}, {Kiekebush}, {Klein}, {Koch}, {Lizon}, {Mehrgan}, {Petr-Gotzens},
  {Pritchard}, {Selman}, \& {Stegmeier}}]{Kissler-Patig:2008fr}
{Kissler-Patig}, M., {Pirard}, J.-F., {Casali}, M., {et~al.} 2008, \aap, 491,
  941

\bibitem[{{Launhardt} {et~al.}(2002){Launhardt}, {Zylka}, \&
  {Mezger}}]{Launhardt:2002nx}
{Launhardt}, R., {Zylka}, R., \& {Mezger}, P.~G. 2002, \aap, 384, 112

\bibitem[{{Martins} {et~al.}(2008){Martins}, {Hillier}, {Paumard},
  {Eisenhauer}, {Ott}, \& {Genzel}}]{Martins:2008aa}
{Martins}, F., {Hillier}, D.~J., {Paumard}, T., {et~al.} 2008, \aap, 478, 219

\bibitem[{{Mauerhan} {et~al.}(2010){Mauerhan}, {Muno}, {Morris}, {Stolovy}, \&
  {Cotera}}]{Mauerhan:2010kb}
{Mauerhan}, J.~C., {Muno}, M.~P., {Morris}, M.~R., {Stolovy}, S.~R., \&
  {Cotera}, A. 2010, \apj, 710, 706

\bibitem[{{Minniti} {et~al.}(2010){Minniti}, {Lucas}, {Emerson}, {Saito},
  {Hempel}, {Pietrukowicz}, {Ahumada}, {Alonso}, {Alonso-Garcia}, {Arias},
  {Bandyopadhyay}, {Barb{\'a}}, {Barbuy}, {Bedin}, {Bica}, {Borissova},
  {Bronfman}, {Carraro}, {Catelan}, {Clari{\'a}}, {Cross}, {de Grijs},
  {D{\'e}k{\'a}ny}, {Drew}, {Fari{\~n}a}, {Feinstein}, {Fern{\'a}ndez
  Laj{\'u}s}, {Gamen}, {Geisler}, {Gieren}, {Goldman}, {Gonzalez}, {Gunthardt},
  {Gurovich}, {Hambly}, {Irwin}, {Ivanov}, {Jord{\'a}n}, {Kerins}, {Kinemuchi},
  {Kurtev}, {L{\'o}pez-Corredoira}, {Maccarone}, {Masetti}, {Merlo},
  {Messineo}, {Mirabel}, {Monaco}, {Morelli}, {Padilla}, {Palma}, {Parisi},
  {Pignata}, {Rejkuba}, {Roman-Lopes}, {Sale}, {Schreiber}, {Schr{\"o}der},
  {Smith}, {}, {Soto}, {Tamura}, {Tappert}, {Thompson}, {Toledo}, {Zoccali}, \&
  {Pietrzynski}}]{Minniti:2010fk}
{Minniti}, D., {Lucas}, P.~W., {Emerson}, J.~P., {et~al.} 2010, \na, 15, 433

\bibitem[{{Morris} \& {Serabyn}(1996)}]{Morris:1996vn}
{Morris}, M. \& {Serabyn}, E. 1996, \araa, 34, 645

\bibitem[{{Nagayama} {et~al.}(2003){Nagayama}, {Nagashima}, {Nakajima},
  {Nagata}, {Sato}, {Nakaya}, {Yamamuro}, {Sugitani}, \&
  {Tamura}}]{Nagayama:2003fk}
{Nagayama}, T., {Nagashima}, C., {Nakajima}, Y., {et~al.} 2003, in \procspie,
  Vol. 4841, Instrument Design and Performance for Optical/Infrared
  Ground-based Telescopes, ed. M.~{Iye} \& A.~F.~M. {Moorwood}, 459--464

\bibitem[{{Najarro} {et~al.}(2009){Najarro}, {Figer}, {Hillier}, {Geballe}, \&
  {Kudritzki}}]{Najarro:2009vn}
{Najarro}, F., {Figer}, D.~F., {Hillier}, D.~J., {Geballe}, T.~R., \&
  {Kudritzki}, R.~P. 2009, \apj, 691, 1816

\bibitem[{{Nataf} {et~al.}(2014){Nataf}, {Cassisi}, \&
  {Athanassoula}}]{Nataf:2014aa}
{Nataf}, D.~M., {Cassisi}, S., \& {Athanassoula}, E. 2014, \mnras, 442, 2075

\bibitem[{{Nishiyama} {et~al.}(2006{\natexlab{a}}){Nishiyama}, {Nagata},
  {Kusakabe}, {Matsunaga}, {Naoi}, {Kato}, {Nagashima}, {Sugitani}, {Tamura},
  {Tanab{\'e}}, \& {Sato}}]{Nishiyama:2006tx}
{Nishiyama}, S., {Nagata}, T., {Kusakabe}, N., {et~al.} 2006{\natexlab{a}},
  \apj, 638, 839

\bibitem[{{Nishiyama} {et~al.}(2006{\natexlab{b}}){Nishiyama}, {Nagata},
  {Sato}, {Kato}, {Nagayama}, {Kusakabe}, {Matsunaga}, {Naoi}, {Sugitani}, \&
  {Tamura}}]{Nishiyama:2006ai}
{Nishiyama}, S., {Nagata}, T., {Sato}, S., {et~al.} 2006{\natexlab{b}}, \apj,
  647, 1093

\bibitem[{{Nishiyama} {et~al.}(2008){Nishiyama}, {Nagata}, {Tamura}, {Kandori},
  {Hatano}, {Sato}, \& {Sugitani}}]{Nishiyama:2008qa}
{Nishiyama}, S., {Nagata}, T., {Tamura}, M., {et~al.} 2008, \apj, 680, 1174

\bibitem[{{Nogueras-Lara} {et~al.}(2018{\natexlab{a}}){Nogueras-Lara},
  {Gallego-Calvente}, {Dong}, {Gallego-Cano}, {Girard}, {Hilker}, {de Zeeuw},
  {Feldmeier-Krause}, {Nishiyama}, {Najarro}, {Neumayer}, \&
  {Sch{\"o}del}}]{Nogueras-Lara:2018aa}
{Nogueras-Lara}, F., {Gallego-Calvente}, A.~T., {Dong}, H., {et~al.}
  2018{\natexlab{a}}, \aap, 610, A83

\bibitem[{{Nogueras-Lara} {et~al.}(2018{\natexlab{b}}){Nogueras-Lara},
  {Sch{\"o}del}, {Dong}, {Najarro}, {Gallego-Calvente}, {Hilker},
  {Gallego-Cano}, {Nishiyama}, {Neumayer}, {Feldmeier-Krause}, {Girard},
  {Cassisi}, \& {Pietrinferni}}]{Nogueras-Lara:2018ab}
{Nogueras-Lara}, F., {Sch{\"o}del}, R., {Dong}, H., {et~al.}
  2018{\natexlab{b}}, \aap, 620, A83

\bibitem[{{Petr} {et~al.}(1998){Petr}, {Coude Du Foresto}, {Beckwith},
  {Richichi}, \& {McCaughrean}}]{Petr:1998vn}
{Petr}, M.~G., {Coude Du Foresto}, V., {Beckwith}, S.~V.~W., {Richichi}, A., \&
  {McCaughrean}, M.~J. 1998, \apj, 500, 825

\bibitem[{{Pfuhl} {et~al.}(2011){Pfuhl}, {Fritz}, {Zilka}, {Maness},
  {Eisenhauer}, {Genzel}, {Gillessen}, {Ott}, {Dodds-Eden}, \&
  {Sternberg}}]{Pfuhl:2011uq}
{Pfuhl}, O., {Fritz}, T.~K., {Zilka}, M., {et~al.} 2011, \apj, 741, 108

\bibitem[{{Portegies Zwart} {et~al.}(2002){Portegies Zwart}, {Makino},
  {McMillan}, \& {Hut}}]{Portegies-Zwart:2002fk}
{Portegies Zwart}, S.~F., {Makino}, J., {McMillan}, S.~L.~W., \& {Hut}, P.
  2002, \apj, 565, 265

\bibitem[{{Primot} {et~al.}(1990){Primot}, {Rousset}, \&
  {Fontanella}}]{Primot:1990fk}
{Primot}, J., {Rousset}, G., \& {Fontanella}, J.~C. 1990, Journal of the
  Optical Society of America A, 7, 1598

\bibitem[{{Reid} \& {Brunthaler}(2004)}]{Reid:2004ph}
{Reid}, M.~J. \& {Brunthaler}, A. 2004, \apj, 616, 872

\bibitem[{{Saito} {et~al.}(2012){Saito}, {Minniti}, {Dias}, {Hempel},
  {Rejkuba}, {Alonso-Garc{\'{\i}}a}, {Barbuy}, {Catelan}, {Emerson},
  {Gonzalez}, {Lucas}, \& {Zoccali}}]{Saito:2012fk}
{Saito}, R.~K., {Minniti}, D., {Dias}, B., {et~al.} 2012, \aap, 544, A147

\bibitem[{{Salaris} {et~al.}(2002){Salaris}, {Cassisi}, \&
  {Weiss}}]{Salaris:2002ab}
{Salaris}, M., {Cassisi}, S., \& {Weiss}, A. 2002, \pasp, 114, 375

\bibitem[{{Sch{\"o}del} {et~al.}(2007){Sch{\"o}del}, {Eckart}, {Alexander},
  {Merritt}, {Genzel}, {Sternberg}, {Meyer}, {Kul}, {Moultaka}, {Ott}, \&
  {Straubmeier}}]{Schodel:2007tw}
{Sch{\"o}del}, R., {Eckart}, A., {Alexander}, T., {et~al.} 2007, \aap, 469, 125

\bibitem[{{Sch{\"o}del} {et~al.}(2014){Sch{\"o}del}, {Feldmeier}, {Kunneriath},
  {Stolovy}, {Neumayer}, {Amaro-Seoane}, \& {Nishiyama}}]{Schodel:2014fk}
{Sch{\"o}del}, R., {Feldmeier}, A., {Kunneriath}, D., {et~al.} 2014, \aap, 566,
  A47

\bibitem[{{Sch{\"o}del} {et~al.}(2018){Sch{\"o}del}, {Gallego-Cano}, {Dong},
  {Nogueras-Lara}, {Gallego-Calvente}, {Amaro-Seoane}, \&
  {Baumgardt}}]{Schodel:2018aa}
{Sch{\"o}del}, R., {Gallego-Cano}, E., {Dong}, H., {et~al.} 2018, \aap, 609,
  A27

\bibitem[{{Sch{\"o}del} {et~al.}(2010){Sch{\"o}del}, {Najarro}, {Muzic}, \&
  {Eckart}}]{Schodel:2010fk}
{Sch{\"o}del}, R., {Najarro}, F., {Muzic}, K., \& {Eckart}, A. 2010, \aap, 511,
  A18+

\bibitem[{{Sch{\"o}del} {et~al.}(2013){Sch{\"o}del}, {Yelda}, {Ghez}, {Girard},
  {Labadie}, {Rebolo}, {P{\'e}rez-Garrido}, \& {Morris}}]{Schodel:2013fk}
{Sch{\"o}del}, R., {Yelda}, S., {Ghez}, A., {et~al.} 2013, \mnras, 429, 1367

\bibitem[{{Scoville} {et~al.}(2003){Scoville}, {Stolovy}, {Rieke},
  {Christopher}, \& {Yusef-Zadeh}}]{Scoville:2003la}
{Scoville}, N.~Z., {Stolovy}, S.~R., {Rieke}, M., {Christopher}, M., \&
  {Yusef-Zadeh}, F. 2003, \apj, 594, 294

\bibitem[{{Yusef-Zadeh} {et~al.}(2009){Yusef-Zadeh}, {Hewitt}, {Arendt},
  {Whitney}, {Rieke}, {Wardle}, {Hinz}, {Stolovy}, {Lang}, {Burton}, \&
  {Ramirez}}]{Yusef-Zadeh:2009ph}
{Yusef-Zadeh}, F., {Hewitt}, J.~W., {Arendt}, R.~G., {et~al.} 2009, \apj, 702,
  178

\end{thebibliography}

\appendix

\section{Tables}

Tables \ref{obs_cen}, \ref{obs_bul}, and \ref{obs_disc} summarise the observing conditions of the data used to produce the GALACTICNUCLEUS survey.

\def\arraystretch{1.25}
\begin{table*}
\begin{center}
\caption{Observing details for the fields used for the central catalogue.}
\label{obs_cen} 
\begin{tabular}{ccccccccccccc}
 &  &  &  &  &  &  &  &  &  &  &  & \tabularnewline
\hline 
\hline 
HAWK-I & Field & Date & Seeing$^a$ & N$^b$ & Field & Date & Seeing$^a$ & N$^b$ & Field & Date & Seeing$^a$ & N$^b$\tabularnewline
filter &  & (d/m/year) & (arcsec) &  &  & (d/m/year) & (arcsec) &  &  & (d/m/year) & (arcsec) & \tabularnewline
\hline 
$J$ &  & 08/06/2015 & 0.37 & 49 &  & 06/06/2015 & 0.58 & 49 &  & 08/06/2015 & 0.40 & 49\tabularnewline
$H$ & F1 & 06/06/2015 & 0.52 & 49 & F2 & 06/06/2015 & 0.70 & 48 & F3 & 06/06/2015 & 0.73 & 49\tabularnewline
$K_s$ &  & 06/06/2015 & 0.57 & 49 &  & 07/06/2015 & 0.60 & 49 &  & 06/06/2015 & 0.60 & 49\tabularnewline
\hline 
$J$ &  & 06/06/2015 & 0.76 & 49 &  & 06/06/2015 & 0.58 & 49 &  & 07/06/2015 & 0.57 & 49\tabularnewline
$H$ & F4 & 06/06/2015 & 0.74 & 49 & F5 & 06/06/2015 & 0.60 & 49 & F6 & 07/06/2015 & 0.63 & 49\tabularnewline
$K_s$ &  & 06/06/2015 & 0.86 & 49 &  & 07/06/2015 & 0.55 & 49 &  & 07/06/2015 & 0.53 & 49\tabularnewline
\hline 
$J$ &  & - & - & - &  & 07/06/2015 & 0.57 & 49 &  & 08/06/2015 & 0.40 & 49\tabularnewline
$H$ & F7 & 07/06/2015 & 0.47 & 50 & F8 & 07/06/2015 & 0.56 & 49 & F9 & 08/06/2015 & 0.40 & 49\tabularnewline
$K_s$ &  & 10/06/2018 & 0.62 & 49 &  & 08/06/2015 & 0.60 & 48 &  & 08/06/2015 & 0.54 & 49\tabularnewline
\hline 
$J$ &  & 08/06/2015 & 0.35 & 49 &  & 08/06/2015 & 0.36 & 49 &  & 08/06/2015 & 0.36 & 49\tabularnewline
$H$ & F10 & 08/06/2015 & 0.46 & 49 & F11 & 09/06/2015 & 0.47 & 49 & F12 & 09/06/2015 & 0.43 & 49\tabularnewline
$K_s$ &  & 08/06/2015 & 0.45 & 49 &  & 09/06/2015 & 0.62 & 49 &  & 09/06/2015 & 0.62 & 49\tabularnewline
\hline 
$J$ &  & 09/06/2015 & 0.46 & 49 &  & 09/06/2015 & 0.40 & 49 &  & 10/06/2015 & 0.53 & 49\tabularnewline
$H$ & F13 & 09/06/2015 & 0.45 & 48 & F14 & 09/06/2015 & 0.53 & 48 & F15 & 10/06/2015 & 0.62 & 49\tabularnewline
$K_s$ &  & 09/06/2015 & 0.48 & 49 &  & 09/06/2015 & 0.66 & 49 &  & 10/06/2015 & 0.64 & 49\tabularnewline
\hline 
$J$ &  & 10/06/2015 & 0.48 & 49 &  & 10/06/2015 & 0.52 & 49 &  & 27/06/2015 & 0.47 & 48\tabularnewline
$H$ & F16 & 10/06/2015 & 0.46 & 47 & F17 & 10/06/2015 & 0.60 & 48 & F18 & 27/06/2015 & 0.32 & 49\tabularnewline
$K_s$ &  & 10/06/2015 & 0.55 & 49 &  & 10/06/2015 & 0.57 & 48 &  & 04/07/2015 & 0.66 & 49\tabularnewline
\hline 
$J$ &  & 27/06/2015 & 0.34 & 49 &  & 27/06/2015 & 0.33 & 49 &  & 27/06/2015 & 0.39 & 48\tabularnewline
$H$ & F19 & 27/06/2015 & 0.38 & 49 & F20 & 21/07/2015 & 0.47 & 49 & F21 & 21/07/2015 & 0.62 & 49\tabularnewline
$K_s$ &  & 02/07/2015 & 0.45 & 49 &  & 05/07/2015 & 0.76 & 48 &  & 12/07/2015 & 0.53 & 49\tabularnewline
\hline 
$J$ &  & 12/07/2015 & 0.42 & 49 &  & 12/07/2015 & 0.41 & 49 &  & 19/07/2015 & 0.38 & 49\tabularnewline
$H$ & F22 & 21/07/2015 & 0.73 & 47 & F23 & 24/07/2015 & 0.41 & 48 & F24 & 25/07/2015 & 0.60 & 49\tabularnewline
$K_s$ &  & 13/07/2015 & 0.56 & 49 &  & 13/07/2015 & 0.51 & 49 &  & 13/07/2015 & 0.68 & 48\tabularnewline
\hline 
$J$ &  & 19/07/2015 & 0.41 & 49 &  & 21/07/2015 & 0.64 & 49 &  & 21/07/2015 & 0.59 & 41\tabularnewline
$H$ & F25 & 24/07/2015 & 0.62 & 46 & F26 & 25/07/2015 & 0.88 & 48 & F27 & 03/10/2015 & 0.51 & 49\tabularnewline
$K_s$ &  & 25/07/2015 & 0.67 & 49 &  & 25/07/2015 & 0.72 & 51 &  & 18/09/2015 & 0.49 & 49\tabularnewline
\hline 
$J$ &  & 20/07/2015 & 0.46 & 49 &  & 23/07/2015 & 0.45 & 49 &  & 24/07/2015 & 0.43 & 49\tabularnewline
$H$ & F28 & 04/10/2015 & 0.50 & 49 & F29 & 27/03/2016 & 0.59 & 50 & F30 & 20/05/2016 & 0.56 & 48\tabularnewline
$K_s$ &  & 27/03/2016 & 0.60 & 30 &  & 12/05/2016 & 0.67 & 47 &  & 07/10/2015 & 0.74 & 48\tabularnewline
\hline 
\end{tabular}
\vspace{0.75cm}

\textbf{Notes.} (a) In-band seeing estimated from the PSF FWHM measured in
long exposure images. (b) Number of pointings. The data corresponding to F7 (J band) were obtained under bad conditions and have not been included in the final data release of the survey.
\end{center}
 \end{table*}

\def\arraystretch{1.25}
\begin{table}
\begin{center}
\caption{Observing details for the fields used for the inner bulge and the transition zone.}
\label{obs_bul} 

\begin{tabular}{ccccc}
 &  &  &  & \tabularnewline
\hline 
\hline 
Filter & Field & Date & Seeing$^a$ & N$^b$\tabularnewline
 &  & (d/m/year) & (arcsec) & \tabularnewline
\hline 
$J$ &  & 24/07/2015 & 0.43 & 49\tabularnewline
$H$ & B1 & 20/05/2016 & 0.56 & 49\tabularnewline
$K_s$ &  & 28/06/2015 & 0.54 & 49\tabularnewline
\hline 
$J$ &  & 24/07/2015 & 0.43 & 49\tabularnewline
$H$ & B2 & 26/05/2016 & 0.33 & 49\tabularnewline
$K_s$ &  & 14/05/2016 & 0.58 & 49\tabularnewline
\hline 
$J$ &  & 24/07/2015 & 0.39 & 51\tabularnewline
$H$ & T3 & 21/05/2016 & 0.51 & 50\tabularnewline
$K_s$ &  & 14/05/2016 & 0.47 & 50\tabularnewline
\hline 
$J$ &  & 27/06/2016 & 0.46 & 49\tabularnewline
$H$ & T4 & 12/06/2016 & 0.79 & 49\tabularnewline
$K_s$ &  & 12/06/2016 & 0.56 & 50\tabularnewline
\hline 
$J$ &  & 27/06/2016 & 0.48 & 49\tabularnewline
$H$ & B5 & 12/06/2016 & 0.83 & 49\tabularnewline
$K_s$ &  & 12/06/2016 & 0.95 & 30\tabularnewline
\hline 
$J$ &  & 27/06/2016 & 0.51 & 49\tabularnewline
$H$ & B6 & - & - & -\tabularnewline
$K_s$ &  & 05/04/2017 & 0.46 & 44\tabularnewline
\hline 
$J$ &  & 27/06/2016 & 0.55 & 49\tabularnewline
$H$ & T7 & 24/04/2017 & 0.42 & 38\tabularnewline
$K_s$ &  & 26/06/2016 & 0.73 & 49\tabularnewline
\hline 
$J$ &  & 27/06/2016 & 0.56 & 49\tabularnewline
$H$ & T8 & 27/06/2016 & 0.54 & 59\tabularnewline
$K_s$ &  & 02/05/2017 & 0.64 & 43\tabularnewline
\hline 
\end{tabular}
\vspace{0.75cm}
\end{center}
\textbf{Notes.} (a) In-band seeing estimated from the PSF FWHM measured in
long exposure images. (b) Number of pointings. The data corresponding to B6 (H band) were obtained under bad conditions and have not been included in the final data release of the survey.

 \end{table}

\def\arraystretch{1.25}
\begin{table*}
\begin{center}
\caption{Observing details for the fields used for the NSD.}
\label{obs_disc} 
\begin{tabular}{ccccccccccccc}
 &  &  &  &  &  &  &  &  &  &  &  & \tabularnewline
\hline 
\hline 
Filter & Field & Date & Seeing$^a$ & N$^b$& Field & Date & Seeing$^a$ & N$^b$ & Field & Date & Seeing$^a$ & N$^b$\tabularnewline
 &  & (d/m/year) & (arcsec) &  &  & (d/m/year) & (arcsec) &  &  & (d/m/year) & (arcsec) & \tabularnewline
\hline 
$J$ &  & 27/06/2016 & 0.49 & 49 &  & 27/06/2016 & 0.81 & 49 &  & 28/06/2016 & 0.87 & 48\tabularnewline
$H$ & D9 & 27/06/2016 & 1.13 & 49 & D10 & 28/06/2016 & 0.91 & 49 & D11 & 28/06/2016 & 0.89 & 49\tabularnewline
$K_s$ &  & 03/05/2017 & 0.77 & 50 &  & 27/06/2016 & 0.81 & 48 &  & 28/06/2016 & 0.90 & 49\tabularnewline
\hline 
$J$ &  & 20/07/2017 & 0.51 & 44 &  & 23/09/2017 & 0.54 & 44 &  & 30/09/2017 & 0.49 & 44\tabularnewline
$H$ & D12 & 03/06/2017 & 0.33 & 44 & D13 & 24/06/2017 & 0.70 & 44 & D14 & 23/07/2017 & 0.66 & 44\tabularnewline
$K_s$ &  & 03/06/2017 & 0.35 & 39 &  & 24/06/2017 & 0.61 & 48 &  & 11/08/2017 & 0.65 & 44\tabularnewline
\hline 
$J$ &  & 01/10/2017 & 0.85 & 44 &  & 27/03/2018 & 0.41 & 44 &  & 21/05/2018 & 0.38 & 44\tabularnewline
$H$ & D15 & 21/09/2017 & 0.75 & 44 & D17 & 22/03/2018 & 0.34 & 43 & D18 & 28/03/2018 & 0.82 & 47\tabularnewline
$K_s$ &  & 11/08/2017 & 0.85 & 44 &  & 27/03/2018 & 0.37 & 43 &  & 28/03/2018 & 0.79 & 44\tabularnewline
\hline 
$J$ &  & 25/05/2018 & 0.49 & 44 &  & 10/06/2018 & 0.88 & 44 &  &  &  & \tabularnewline
$H$ & D19 & 21/05/2018 & 0.41 & 44 & D21 & 25/05/2018 & 0.59 & 44 &  &  &  & \tabularnewline
$K_s$ &  & 16/04/2018 & 0.54 & 44 &  & 06/06/2018 & 0.51 & 44 &  &  &  & \tabularnewline
\hline 
\end{tabular}
\vspace{0.75cm}

\textbf{Notes.} (a) In-band seeing estimated from the PSF FWHM measured in
long exposure images. (b) Number of pointings.

\end{center}
 \end{table*}

\end{document}